\documentclass[12pt,preprint]{aastex}
\bibpunct{(}{)}{;}{a}{}{;}

\shorttitle{Simulations of Convection in Young Suns}

\newcommand{\vect}[1]{\mbox{\boldmath$#1$\unboldmath}}
\newcommand{\unit}[1]{\ensuremath{\:\mathrm{#1}}}

\newcommand{\DD}{\vect{\mathcal{D}}}
\newcommand{\nab}{\vect{\nabla}}

\newcommand{\rb}{\bar{\rho}}
\newcommand{\Tb}{\bar{T}}
\newcommand{\Sbb}{\bar{S}}
\newcommand{\Pb}{\bar{P}}

\newcommand{\Vp}{v_{\phi}}
\newcommand{\Vt}{v_{\theta}}
\newcommand{\Vrh}{\hat{v}_r} 
\newcommand{\Vth}{\hat{v}_{\theta}}
\newcommand{\Vph}{\hat{v}_{\phi}}

\newcommand{\Vrr}{\tilde{v}_r} 
\newcommand{\Vtr}{\tilde{v}_{\theta}}
\newcommand{\Vpr}{\tilde{v}_{\phi}}
\newcommand{\Vvr}{\tilde{v}}

\newcommand{\BCZ}{\textsc{bcz}}
\newcommand{\PMS}{\textsc{pms}}
\newcommand{\ZAMS}{\textsc{zams}}

\begin{document}


\title{Simulations of turbulent convection in rotating young solar-like stars:\\
Differential rotation and meridional circulation}


\author{J{\'e}r{\^o}me Ballot\altaffilmark{1}, Allan Sacha Brun\altaffilmark{2}, Sylvaine Turck-Chi{\`e}ze\altaffilmark{2}}
\affil{DSM/DAPNIA/Service d'Astrophysique, CEA/Saclay, 91191 Gif-sur-Yvette {\sc cedex}, France}
\email{jballot@mpa-garching.mpg.de, allan-sacha.brun@cea.fr, sylvaine.turck-chieze@cea.fr}
\altaffiltext{1}{Max-Planck-Institut f{\"{u}}r Astrophysik, Karl-Schwarzschild-Str. 1, 85748 Garching, Germany}
\altaffiltext{2}{AIM -- Unit\'e mixte de recherche CEA - CNRS - Universit\'e Paris VII -- 
   UMR n$^\mathrm{o}$ 7158, CEA/Saclay, 91191 Gif-sur-Yvette {\sc cedex}, France}

\begin{abstract}
We present the results of three-dimensional simulations of the deep convective envelope of
a young (10 Myr) one-solar-mass star, obtained with the Anelastic Spherical Harmonic code.
Since young stars are known to be faster rotators than their main sequence counterparts,
we have systematically studied the impact of the stellar rotation speed, by considering stars spinning up to five times as fast as the Sun.
The aim of these nonlinear models is to
understand the complex interactions between convection and rotation.
We discuss the influence of the turbulence level and of the rotation rate on the intensity and the topology of the mean flows.
For all of the computed models, we find a solar-type superficial differential rotation, with an equatorial acceleration, and meridional circulation that exhibits a multicellular structure. Even if the differential rotation contrast $\Delta\Omega$ decreases only marginally for high rotation rates, the meridional circulation intensity clearly weakens according to our simulations.
We have also shown that, for Taylor numbers above a certain threshold ($T_a\gtrsim 10^9$), the
convection can develop a vacillating behavior. Since simulations with high turbulence levels and rotation rates exhibit strongly cylindrical internal rotation profiles, we have considered the influence of
baroclinic effects at the base of the convective
envelope of these young Suns, to see whether such effect can modify the otherwise near cylindrical profiles to produce more conical, solar-like profiles.
\end{abstract}


\section{Introduction}\label{Sec:intro}

Young stars are the subjects of intense observational and
theoretical research.
There is a diversity of interesting objects, from massive Herbig stars to low-mass
T Tauri stars.
We focus in this paper on low-mass stars.
During the pre-main sequence (\PMS) they evolve from the birthline along the Hayashi track; during this phase they are fully convective.
Next a radiative core develops and grows until the stars
reach the zero-age main sequence ({\ZAMS}) and start burning Hydrogen. Thus
energy transfer by convection takes place in a large portion of the {\PMS} stars and, as such, 
is a key process for understanding their structure and their evolution.

During their youth, these stars are surrounded by a disk, more or less dense and thick according to their evolution stage, with possible accretion from the disk to the star. Stars and disks are
generally magnetically coupled. Because of the nature of these interactions, there is a large range of stellar rotation speed. There are slow rotators --- spinning at the solar rate --- and stars spinning a hundred times faster. As a direct consequence, stars on the {\ZAMS} exhibit a very large dispersion of equatorial velocities, but T~Tauri stars present more moderate rotation periods (typically  between 2 and 15 days). The velocity distribution can be well modeled by considering the interaction between the star and its surrounding disk \citep{BouvierF97}. 
The rotation speed for a star on the {\ZAMS} directly depends on the duration over which the central body and the circumstellar material have been coupled in earlier phases.
During the main sequence, because of angular momentum losses through a stellar wind, stars slow down and their rotation speeds follow Skumanich's law \citep{Skumanich72}.
Thus young stars are generally faster rotators than their more evolved brethren.
Such higher rotation rates should impact the internal dynamics of the stars and this impact must be studied with care.
In order to progress on our understanding of such objects, one has to rely on both precise modeling of their internal structure and dynamics, and accurate observations.

Three-dimensional hydrodynamic simulations are useful tools to understand
complex and intricate physical processes, like the interaction between turbulent convection,
fast rotation and possibly magnetic fields.
Successes of such simulations in reproducing solar observations, especially internal
constraints on rotation provided by helioseismology \citep{MieschE00,ElliottM00,BrunT02,BrunM04,MieschB06}, encourage us to pursue this effort in modeling
other stars. 3-D simulations of convective cores of A-type stars have thus been performed. The properties of convection  and angular momentum redistribution were analyzed in purely hydrodynamical models first \citep*{BrowningB04}, then dynamo processes were studied with MHD simulations \citep*{BrunB05}. Simulations of fully convective stars, like T~Tauri ones, have also been performed by
\citet*{DoblerS06}. They have performed 3-D MHD simulations of a convective sphere embedded in a Cartesian box, aiming to model dynamo processes in fully convective stars.
In this paper, our main objective is to continue this effort by performing 3-D hydrodynamical simulations of young solar-like stars.

Today's observations are mostly limited to the stellar surface, even though with the recent launch of CoRoT, deep probing of the internal structure will be soon accessible.
Despite being currently unable to reach the accuracy of observations of the solar interior given by helioseismic inversions,
we can expect that asteroseismology will provide constraints on physical processes, such as the
convection in the envelopes of stars \citep*{Monteiro00,MazumdarA01,RoxVor01,Ballot04BCZ}, especially the young stars in open clusters \citep*{PiauB05}.
Among the current available proxies used to infer the inner part of young stars, Lithium is playing a central role. Lithium is burned by proton capture at temperature around 2--3$\times 10^6$~K. Thus, its surface abundance is very sensitive to the temperature at the base of the convective zone ({\BCZ}) and the physical processes --- diffusion, mixing --- occurring in this region \citep*{BrunT99}.
It provides an efficient mean for checking deficiencies in classical modeling of {\PMS}-stellar structure.
Indeed the evolution of $^7$Li abundance during the {\PMS} is not so easy to predict. \citet{PiauTC02} have shown that $^7$Li burning depends strongly on many ingredients during its short phase of burning --- typically from 2 to 10 Myr for young suns --- which corresponds to the contraction phase and a large increase of the temperature and density at the {\BCZ}. Models of solar-like stars show a stronger depletion of Lithium than the observed ones \citep{VenturaZ98,PiauTC02}, with a great dependence of the internal composition \citep[see also][]{SestitoD06} and of dynamical processes. The convective zone is clearly a crucial element and many questions have been addressed on the role of magnetic field on its extension and also of the effects of turbulence in this region. In this context, it is time to perform 3-D simulations of young stars, and see what we can learn from these models, first when convection and rotation interact, then when magnetic field is also considered (upcoming paper). 

In order to constrain these 3-D simulations, we summarize what is our current knowledge on their dynamics. The rate of differential rotation in latitude is known for many stars. Several techniques exist to infer this differential rotation; they mainly rely on the stellar magnetic activity.
The first method is to use the migration during activity cycles of spots which rise through the
surface at higher latitudes at the beginning of a cycle than at the end, according to the solar paradigm.
Observations of luminosity fluctuations indicate the rotation period for the mean latitude where spots emerge. Thus studying the period modulation within cycles gives a measurement of differential rotation. 
Long term photometric monitoring by \citet{HenryE95} revealed such secular modulation in period signature. They deduced that the contrast of differential rotation $\Delta\Omega$ is almost independent of rotation ($\Delta\Omega\propto\Omega^{0.24\pm0.06}$).
\citet{MessinaG03} have also deduced from long-term photometry the differential rotation rate for few young solar-like stars and found a correlation $\Delta\Omega\propto\Omega^{0.58\pm0.5}$ that is marginally significant relative to the error bars.
Using Mount Wilson Survey of chromospheric \ion{Ca}{2} emission-line fluxes
for an hundred of stars,
\citet*{DonahueS96} have found a strong dependence $\Delta\Omega\propto\Omega^{0.7\pm0.1}$.
By using another method which consists of looking for differential rotation as variations of stellar absorption lines, \citet{ReinersS03a} also have obtained a positive correlation $\Delta\Omega\propto\Omega^{0.53\pm0.35}$.

However, by focusing on young solar-type stars, \citet{CameronB01} have found a relation
$\Delta\Omega\sim{}$constant. This relation is obtained by processing separately K-type and G-type stars. $\Delta\Omega$ does not vary significantly inside each class but is larger for G stars than for K ones.
A more systematic study of \citet{BarnesC05}, indicates such a strong dependence on effective
temperature but a marginal sensitivity to rotation ($\Delta\Omega\propto\Omega^{0.15\pm0.10}$). This study is mainly based on another, more direct, technique, which consists of following displacements of spots on the stellar surface using Doppler imaging \citep*{PetitD02}.

Up to now, theoretical predictions of differential rotation in convective envelopes are mainly produced from mean field models \citep{KitchatinovR95,KitchatinovR99,KukerR97,RudigervR98,KukerS01,Rempel05a,Rempel05b}.

In this paper we aim to analyze the dynamics in the thick convective shell of a young (10~Myr)
solar-type star, and particularly the influence of rotation on the convective motions
and resulting mean flows --- differential rotation, meridional circulation --- in such a star, in order to provide scaling laws useful for observers and 1-D stellar evolutionary models.
After defining the equations solved by the anelastic spherical harmonic (ASH) code and giving specifications of our different models (\S~\ref{Sec:posing}), we describe the spatial and temporal properties of convection in these simulations (\S~\ref{sec:conv}).
Differential rotation profiles and their evolution with $\Omega$ is discussed in \S~\ref{sec:DR} while \S~\ref{sec:MC} is dedicated to the meridional circulation. The physical interpretation of all these results is discussed in \S~\ref{sec:interp}.
Section~\ref{Sec:OscConv} is focused on a peculiar form of vacillating convection we
have found during this study. In conclusion (\S~\ref{sec:concl}) main results are summarized and perspectives are proposed.

\section{Posing the Problem}\label{Sec:posing}

\subsection{Anelastic equations}
The simulations described here were performed using the Anelastic Spherical
Harmonic (ASH) code.  ASH solves the three-dimensional anelastic equations
of motion in a rotating spherical geometry using a pseudo-spectral
semi-implicit approach \citep{CluneE99, MieschE00,BrunM04}.
These equations are fully nonlinear in velocity variables and linearized in
thermodynamic variables with respect to a spherically symmetric mean state.
This mean state is taken to have density $\bar{\rho}$, pressure $\bar{P}$,
temperature $\bar{T}$, specific entropy $\bar{S}$; perturbations about this
mean state are written as $\rho$, $P$, $T$, and $S$.  Conservation of mass,
momentum, and energy in this rotating reference frame are therefore written as
\begin{equation}
\nab \cdot (\rb\vect{v}) = 0,
\end{equation}
\begin{equation}
\begin{array}{r}
\rb \left(\frac{\partial \vect{v}}{\partial t} + (\vect{v}\cdot\nab)\vect{v} + 2\vect{\Omega_o}\times\vect{v}\right)
 = -\nab P\\ 
+ \rho \vect{g} - \nab\cdot\DD -[\nab\bar{P}-\rb\vect{g}],
\end{array}
\end{equation}
\begin{equation}
\begin{array}{r}
\rb \Tb \frac{\partial S}{\partial t}
=\nab\cdot[\kappa_r \rb c_P \nab
(\Tb+T)+\kappa \rb \Tb \nab (\Sbb+S)] \\ 
-\rb \Tb\vect{v}\cdot\nab (\Sbb+S) \\
+2\rb\nu\left[e_{ij}e_{ij}-1/3(\nab\cdot\vect{v})^2\right],
\end{array}
\end{equation}
where $c_P$ is the specific heat at constant pressure, $\vect{v}=(v_r,v_{\theta},v_{\phi})$ is the local velocity in spherical
geometry in the rotating frame of constant angular velocity $\vect{\Omega_o}$, $\vect{g}$ is the gravitational acceleration, $\kappa_r$ is the
radiative diffusivity, and $\DD$ is the viscous stress tensor,
with components
\begin{eqnarray}
{\cal D}_{ij}=-2\rb\nu[e_{ij}-1/3(\nab\cdot\vect{v})\delta_{ij}],
\end{eqnarray}
where $e_{ij}$ is the strain rate tensor.  Here $\nu$ and $\kappa$ are
effective
eddy diffusivities for vorticity and entropy.  To close the set of equations,
linearized relations for the thermodynamic fluctuations are taken as
\begin{equation}
\frac{\rho}{\rb}=\frac{P}{\bar{P}}-\frac{T}{\Tb}=\frac{P}{\gamma\bar{P}}
-\frac{S}{c_P},
\end{equation}
assuming the ideal gas law
\begin{equation}\label{eq:GP}
\bar{P}={\cal R} \rb \Tb ,
\end{equation}
where ${\cal R}$ is the gas constant.  The effects of
compressibility on the convection are taken into account by means of the
anelastic approximation, which filters out sound waves that would otherwise
severely limit the time steps allowed by the simulation.

Convection in stellar environments occurs over a large range of scales.
Numerical simulations cannot, with present computing technology, consider
all these scales simultaneously.  We therefore seek to resolve the largest
scales of the nonlinear flow, which we think are likely to be the dominant
players in establishing differential rotation and other mean properties of
the convection zone.  We do so within a large-eddy simulation
formulation, which explicitly follows larger scale flows while employing
subgrid-scale descriptions for the effects of the unresolved
motions.  Here, those unresolved motions are treated as enhancements to the
viscosity and thermal diffusivity ($\nu$ and $\kappa$), which are thus
effective eddy viscosities and diffusivities.  For simplicity, we have
taken these to be functions of radius alone, and to scale as the inverse of
the square root of the mean density.  We emphasize that currently tractable 
simulations are still many orders of magnitude away in parameter space from the highly
turbulent conditions likely to be found in real stellar convection zones.
These large-eddy simulations should therefore be viewed only as indicators
of the properties of the real flows.  We are encouraged, however, by the
success that similar simulations (cf. \S~\ref{Sec:intro}) have enjoyed 
in matching the detailed observational constraints for the differential 
rotation within the solar convection zone provided by helioseismology.

\subsection{Numerical approach}
Velocity and thermodynamic variables within ASH are expanded in spherical harmonics
$Y^m_{\ell}(\theta,\phi)$ in the horizontal directions and in Chebyshev
polynomials $T_n (r)$ in the radial.  Spatial resolution is thus uniform
everywhere on a sphere when a complete set of spherical harmonics of degree
$\ell$ is used, retaining all azimuthal orders $m$.  We truncate our
expansion at degree $\ell=\ell_\mathrm{max}$, which is related to the number of
latitudinal mesh points $N_{\theta}$ as $\ell_\mathrm{max}=(2N_{\theta}-1)/3$,
take $N_{\phi}=2 N_{\theta}$ latitudinal mesh points, and utilize $N_r$
collocation points for the projection onto the Chebyshev polynomials. 
The grid resolution $N_r \times N_{\theta} \times N_{\phi}$ (cf. Table~\ref{Tab:descrip}) we have
considered depends on the degree of turbulence of the model.
An implicit, second-order Crank-Nicholson
scheme is used in determining the time evolution of the linear terms,
whereas an explicit second-order Adams-Bashforth scheme is employed for the
advective and Coriolis terms.  The ASH code has been optimized to run
efficiently on parallel supercomputers such as the HP ES45, on which the
simulations of this work have been performed.

\subsection{Setting the young sun model}

We describe and analyze in this paper a simplified 3D model of the
convective envelope of a young one-solar-mass star where the plasma is composed only of
a mixture of fully ionized Hydrogen and Helium. This model is built from the structure of an
accurate 1-D stellar model (model Y) to get realistic values for the radiative opacity, density, and
pressure profiles.

Model Y is obtained with the CESAM stellar
evolution code \citep{Morel97} for an age of 10 Myr from the birthline. At this age, the star is located at the turning point  of the {\PMS} evolutionary track, just after the Hayashi
descent. Its radiative core is developing quickly and there is a strong lithium burning \citep[][Fig. 1]{PiauTC02}.
It is characterized by an effective temperature of $4590\unit{K}$, a
luminosity  $L_*=1.86\times10^{33}\unit{erg\,s^{-1}}$  ($0.48\unit{L_\sun}$) and a
radius $R_*=7.67\times10^{10}\unit{cm}$  ($1.1\unit{R_\sun}$).
The outer convective zone covers 45\% of the stellar radius or 37\% of the total mass. Such a convective envelope is almost
twice as thick as the solar one, and, more importantly, 15 times more massive. For such a model, we use the OPAL
equation of state \citep{EOS_OPAL96,EOS_OPAL2001a} 
and the nuclear cross sections of \citet{Adelberger98} to describe the microscopic properties of the stellar plasma.
OPAL opacities \citep{OPAL96} are completed at low temperature by
tables of \citet{Alexander94}.
Convection is computed using the classical mixing length
treatment calibrated to match the observations of the Sun at 4.6~Gyr ($\alpha\approx1.9$).

\begin{figure}
\epsscale{1}
\plotone{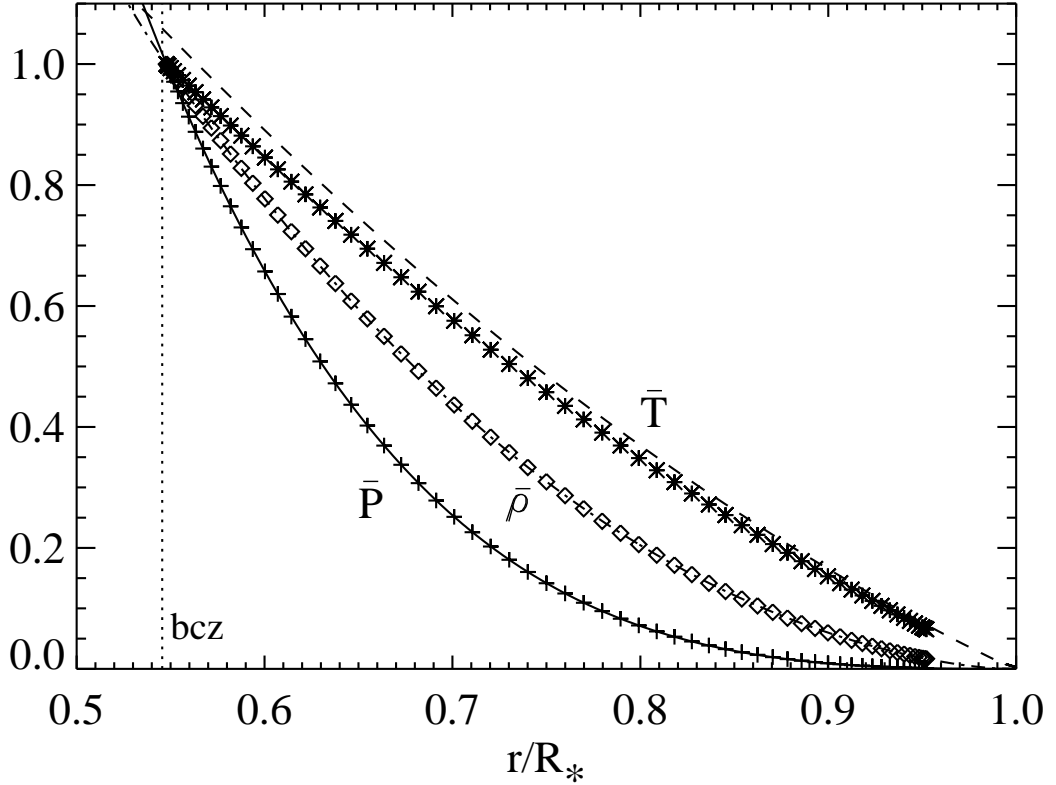}
\figcaption{Thermodynamic mean radial stratification in the outer part of a young 1-M${}_\sun$ star,
within both the 1-D stellar model Y and 3-D hydrodynamic case Yb1. The mean temperature $\Tb$, density $\rb$ and pressure
$\Pb$ are plotted as curves for the model Y and with symbols at their mesh locations for our case Yb1.
The base of the outer convective zone (bcz) is indicated by the dotted line. Variables are normalized to the value they have in model Yb1 at the bottom of the computational domain: $\Tb_{b}=3.4\times10^6\unit{K}$, $\rb_{b}=1.9\unit{g\,cm^{-3}}$, and
$\Pb_{b}=9.1\times10^{14}\unit{dyne\,cm^{-2}}$.\label{fig:profil1D}}
\end{figure}

The  simulations performed with the ASH code were initialized from this model Y. 
The gravity $g$, radiative diffusivity
$\kappa_{r}$, and mean density $\bar{\rho}$ radial profiles
are the starting points for an iterative Newton-Raphson
solution of the hydrostatic balance and for determining the gradients of the
thermodynamic variables.  The mean temperature $\bar{T}$ is then deduced
from equation \ref{eq:GP}.  This technique yields mean profiles in good
agreement with the initial 1-D stellar model Y (Fig.~\ref{fig:profil1D}).

The computational domain extends from about 0.55 to 0.95$\:R_*$.
The overall density contrast in radius is around 60, implying to noticeable
compressibility effects.
The simulations do not model the outer most layers of the star, where the convective scales become very small due to the sudden decrease of the density scale height. This choice excludes the H and He recombination zone and the superadiabatic layer for which the microscopic description is more complex.
The inner boundary corresponds to the base of the convective zone;
the presence of a potential overshooting layer is not taken into account in this work.

We impose impenetrable and stress-free conditions
for the velocity field and a constant flux --- i.e., a constant entropy gradient ---
at both inner and outer boundaries, except for the model Yc5S discussed in
\S~\ref{Sec:OscConv}.  In this model, an entropy gradient in latitude is imposed at the bottom of the shell to model the influence of a tachocline as in \citet*{MieschB06}.

\subsection{Early phases and relaxation of 3-D models}\label{ssec:1stevol}
\begin{figure}
\epsscale{1.0}
\plotone{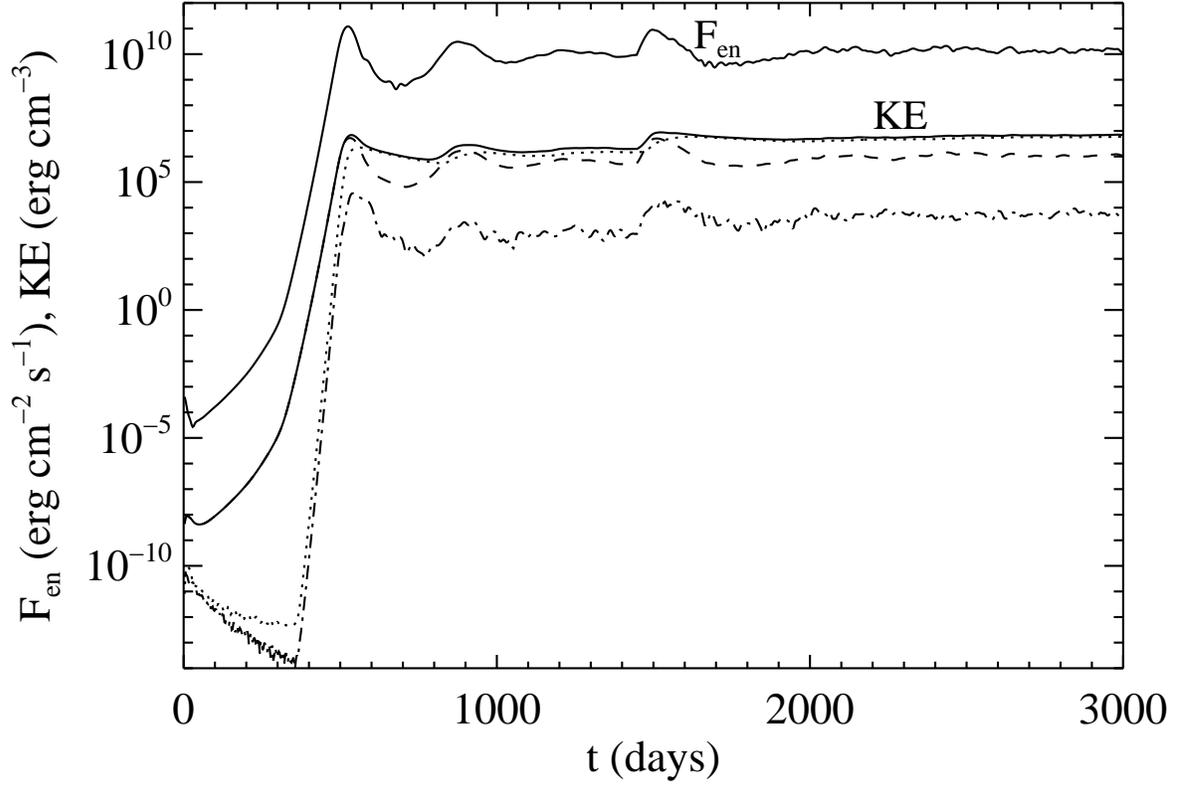}
\figcaption{Evolution of the enthalpy flux $F_{en}$ and of the kinetic energy density (KE), averaged over the full domain from the quiescent state to a statistically stationary state ({\itshape solid lines}). The evolution of the three components of KE are also plotted: the kinetic energy in convective motions ({\itshape dashed line}), in differential rotation ({\itshape dotted line}), and in meridional circulation ({\itshape dash-dotted line}).\label{Fig:evolFen}}
\end{figure}

Some simulations have been started from a quiescent
state with a uniform rotation; others use evolved solutions in which we
modify diffusivities or rotation rate.
One can track the development of the convection by following the mean enthalpy flux crossing the shell. The enthalpy flux is defined as
\begin{equation}\label{eq:Fen}
F_{en}=v_r \rb c_P T
\end{equation}
Figure~\ref{Fig:evolFen} shows the temporal evolution of $F_{en}$, averaged over the full domain, along with the evolution of mean kinetic energy KE, and its axisymmetric and non-axisymmetric components (see \S~\ref{ssec:energy} for more details).
Given the high Rayleigh numbers used in all our models, convective instabilities develop from the quiescent state and grow exponentially over about 500 days, until saturation. After a phase of $\sim$1500 days over which relaxation oscillations occur, statistically stationary states are usually reached. Only statistical fluctuations modulate the temporal trace in the late evolution. When such a state has been attained, we have integrated the model about a thousand days, namely several convective-overturning times and several rotation periods, even for the most consuming simulations, for a total of one-half million node hours spent.

\begin{figure}
\epsscale{1.0}
\plotone{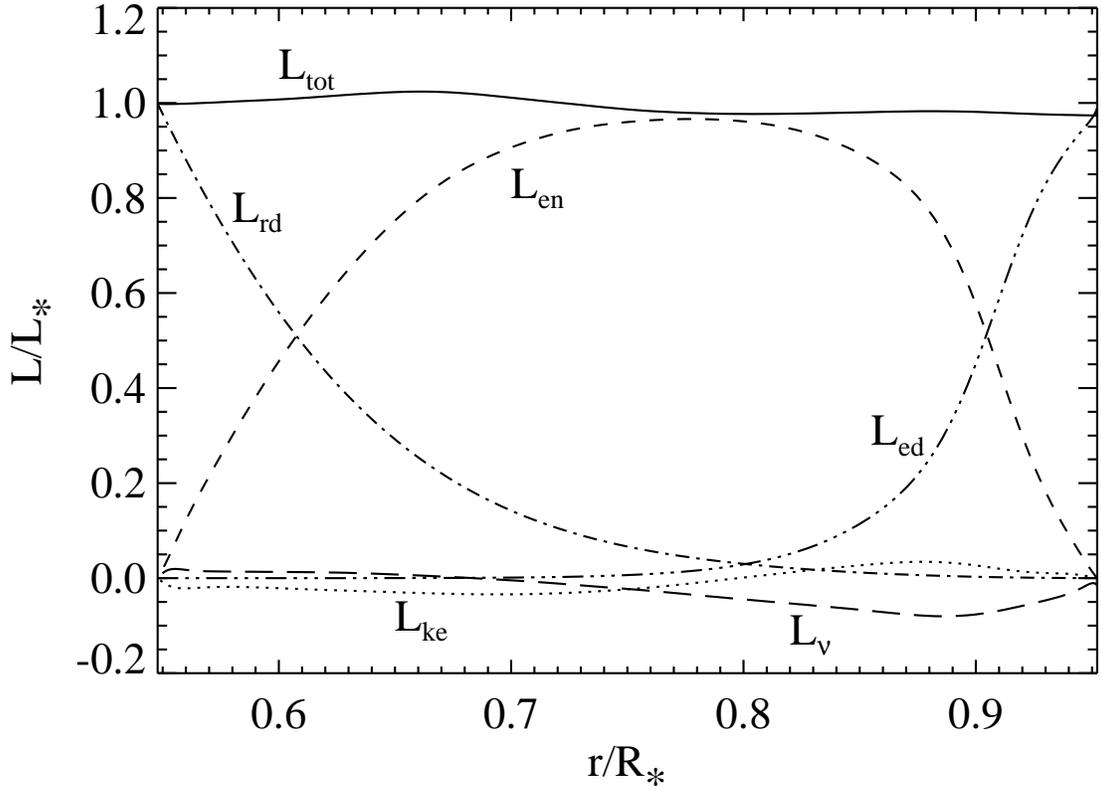}
\figcaption{Radial transport of energy in case Yc1 achieved by the fluxes $L_{en}$, $L_{rd}$, $L_{ke}$, $L_{ed}$, $L_{\nu}$, and their total $L_{tot}$, all normalized by the stellar luminosity $L_*$. \label{Fig:flux}}
\end{figure}

One can see for example how the energy is transfered from the bottom to the top of the shell, on average, for a relaxed simulation.
Several processes take part in this transfer \citep[for detailed expressions, see equations {[11]}-{[16]} in][]{BrunM04}.
Figure~\ref{Fig:flux} shows the horizontally averaged distribution of each flux in a typical model (Yc1),
after averaging over time (1200~days). 
The main component is of course the enthalpy luminosity $L_{en}$ advected by convection (luminosities $L$
are defined from fluxes $F$ by the relation $L=4\pi r^2 F$).
Close to the bottom, radiative luminosity $L_{rd}$ becomes significant because of the steady increase of radiative conductivity $\kappa_r$ with depth.
Near the top of the layer, the heat transport is dominated by the subgrid-scale turbulence that yields the term $L_{ed}$. This flux is proportional to a specified eddy diffusivity and to the mean radial gradient of entropy and serves to carry the total flux through the upper boundary and prevents the entropy gradient there from becoming superadiabatic compared to the scales of convection that we are able to resolve spatially.
There are two minor extra contributions. The first is the kinetic energy flux ($F_{ke}$), which is slightly
negative, except when $L_{ed}$ dominates. This negative value is a consequence of the fast downflow sheets and plumes achieved by effects of compressibility \citep{HurlburtT86}.
The second is the viscous flux $L_\nu$, which contributes to the total flux mainly near the top where there is a viscous layer.
Profiles are very similar in the other simulations.

In this work we will concentrate on three series of models, respectively called  Ya, Yb and Yc. These models possess different levels of turbulence and Prandtl numbers 
$P_r$=4, 1 or $1/4$ ($P_r=\nu/\kappa$).
For each Prandtl number considered, we have performed models with 
$\Omega_o=1$, 2 and $5\,\Omega_\sun$ ($\Omega_\sun=2.6\times 10^{-6}\unit{rad\,s^{-1}}$,
 corresponds to a period of 28 days), for a total of nine models.
We have also computed five supplementary models
spinning at $5\,\Omega_\sun$, with higher turbulence levels, different Prandtl numbers or
boundary conditions. This study spans a large range of Taylor and Rayleigh numbers,
$T_a \in [10^6,10^{10}]$ and $R_a \in [10^5,10^7]$. All the model parameters and characteristics are detailed in Table~\ref{Tab:descrip}.

\section{Convective patterns and their evolution in deep spherical shell}\label{sec:conv}

Convection in young solar stars is a major player since it transports heat outward from the deep 
internal parts, redistributes angular momentum and establishes large scale mean flows.
Understanding these intricate physical processes is thus crucial if we want to progress
in our knowledge of such stars. 
We here start by describing in details the convective patterns established in our simulations,
how they are modified by the rotation rate and the turbulence level and how these 
nonlinear interactions lead to complex temporal evolutions.

\subsection{Topology of the convection}

\begin{figure*}
\epsscale{1.0}
\plotone{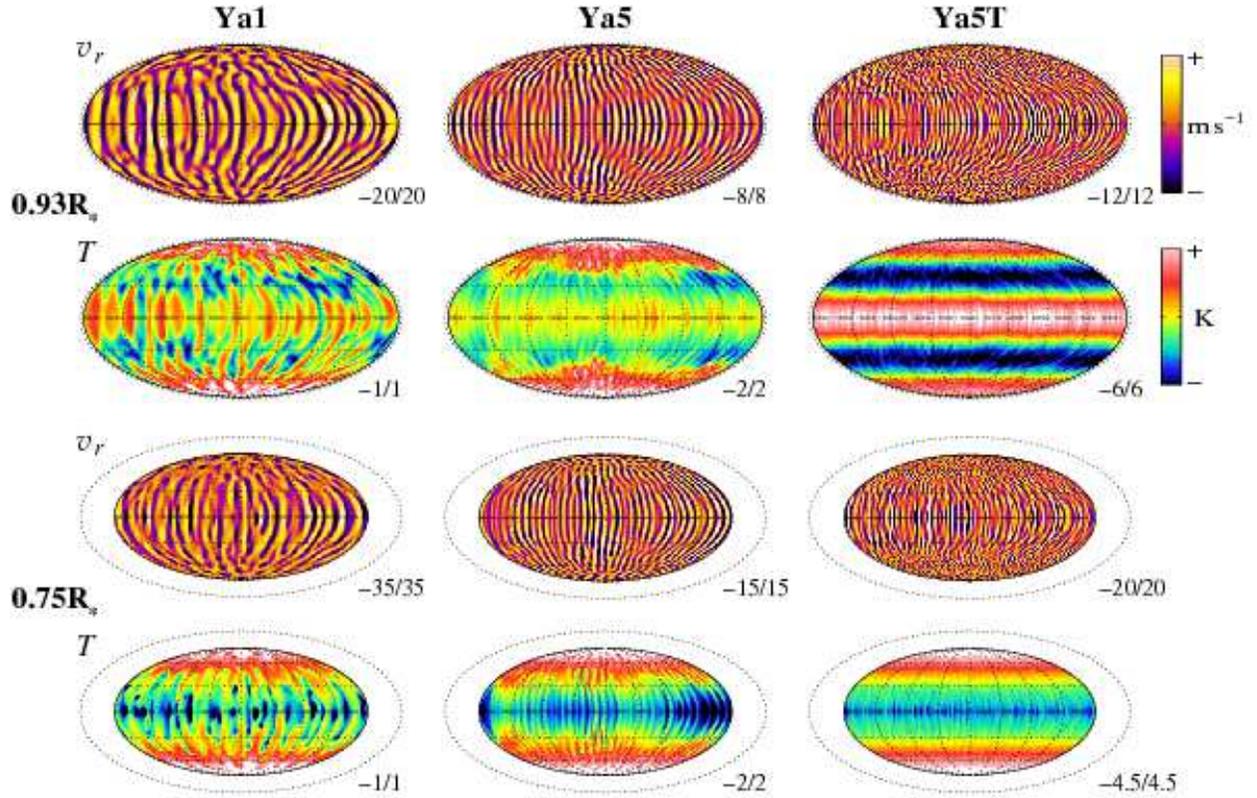}
\figcaption{Radial velocity $v_r$ and temperature fluctuation $T$ fields near the surface ($0.93\:R_*$) and at mid-depth ($0.75\:R_*$), at a given instant for the three models Ya1, Ya5, and Ya5T. Maps are shown in Mollweide projection. Dashed horizontal line designates equator. Meridians are plotted with dotted lines every 45\degr\ and parallels every 30\degr. Dotted ellipse shows the relative position of the stellar surface. Minimum and maximum amplitudes for each field are indicated next to each panel.\label{Fig:shsl_Ya}}
\end{figure*}

Figure~\ref{Fig:shsl_Ya} shows maps of radial velocity $v_r$ and temperature fluctuation $T$ near the surface and at mid-depth of the convective envelope for models of the Ya series.
Near the surface, the velocity field of the Ya1 simulation exhibits networks of downward flows confined to the periphery of convective cells, and broader and weaker upward flows in their centers. This asymmetry is made possible by the anelastic approach which includes compressibility effects. This quite laminar model has a ``classic'' structure composed, in the equatorial area, of so-called banana cells, aligned with rotation axis. It is a well-known pattern in solar modeling \citep[e.g.,][]{MieschE00}. Around the poles, convective cells look less elongated.

By comparing maps at different depths, one recovers some patterns at similar places, which shows the great connectivity of the downflow network across the whole convective domain. In fact, upward and downward flows are not radial, i.e. aligned with the local gravity, but are tilted by the rotation and tend to be aligned with the polar axis. In Ya1 model, upward structures can start, in a meridional plane, at the equator at the bottom of the shell and reach the top near a latitude of 30\degr.
Around the poles, tilts of structures are less significant because $\vect{\Omega_o}$ are $\vect{g}$ are almost aligned.

Comparing radial velocity $v_r$ and temperature fluctuation $T$ maps shows their strong correlation. It just means that hotter material goes up and cooler material sinks. This strong correlation of $v_r$ and $T$ leads to large outward enthalpy flux and implies heat transport by convection (see Fig.~\ref{Fig:flux}).

Temperature maps show also a clear variation with latitude. This zonal structure varies with depth. At mid-depth (as at the bottom, not shown), there is a monotonic variation from the hotter poles to the cooler equator. Close to the surface, the temperature increases again in the equatorial region. The amplitude of these variations is around 2\unit{K}.
This axisymmetric thermal variation between poles and equator plays an important role in the understanding of differential rotation profiles (see discussion \S~\ref{sec:interp}).

When the rotation speed is increased (see case Ya5), the pattern shape is modified but retains its overall organization.
Asymmetry of flow is reduced because the convection is less vigorous. 
Indeed, we notice that speeds of convective flows have been significantly reduced, because
rotation tends to stabilize flows \citep[e.g.,][]{Chandrasekhar}.
Stronger spatial variations in flow velocities also appear, which can be seen on the $v_r$ maps as variations on the image contrast.
Another noticeable difference is the pattern size. In the Ya5 model, convective patterns are clearly smaller than in Ya1. This is also a well-known effect, which can be understood simply by following the linear growth of convective instabilities: degrees $\ell_c$ of the most instable modes increase with the Taylor number \citep{GlatzGil81b}. From the maps, we can estimate that $\ell_c\sim18$ in Ya1 and increases to 42 in Ya5. We have performed an extra model thus Ya5T, similar to Ya5 but more turbulent of which $T_a$ is 16 time higher; in this case, estimated $\ell_c$ goes up to 60, producing even finer structures.
In addition, an important impact of rotation on flows is the tilts of structures are strongly pronounced in the equatorial area. Convective structures are essentially aligned with the rotation axis in the Ya5 case (and somewhat less in Ya5T). In this case, the Coriolis term dominates the others; convection tends to develop along Busse's columns as a consequence \citep[see for example][]{Busse02}.

Finally, the rotation does not affect the structure of the temperature fluctuations. The zonal structure of $T$ is very similar to those of Ya1 in that, the same kind of bands are visible; only the thermal contrasts have been reinforced, especially for Ya5T, in agreement with the stronger zonal flow present in this case (see \S~\ref{sec:DR} and \ref{sec:interp}).

Convective motions obviously modify the radial structure. In a deep convective zone, the radial entropy gradient is negative (unstable region), but close to zero, especially at the bottom of the shell where the convection layer is close to adiabatic.
Table~\ref{Tab:descrip} shows that for models with exactly the same diffusivities $\nu$ and $\kappa$, the Rayleigh number increases with rotation rate. It indicates that rotation modifies $d\Sbb/dr$; that is confirmed by Fig.~\ref{Fig:dsdr_Ya}, which shows the entropy gradient for the Ya model series.
As described above for the velocity fields, the stabilizing effects of rotation reduce the vigor of the flow. As convection is less vigorous, it becomes less efficient and the entropy gradient needs to become steeper to evacuate energy. For the same reason, since the intensity of convection is greater in the simulation Ya5T, more turbulent than Ya5, the entropy gradient is less strong in Ya5T.

\begin{figure*}
\epsscale{1.0}
\plotone{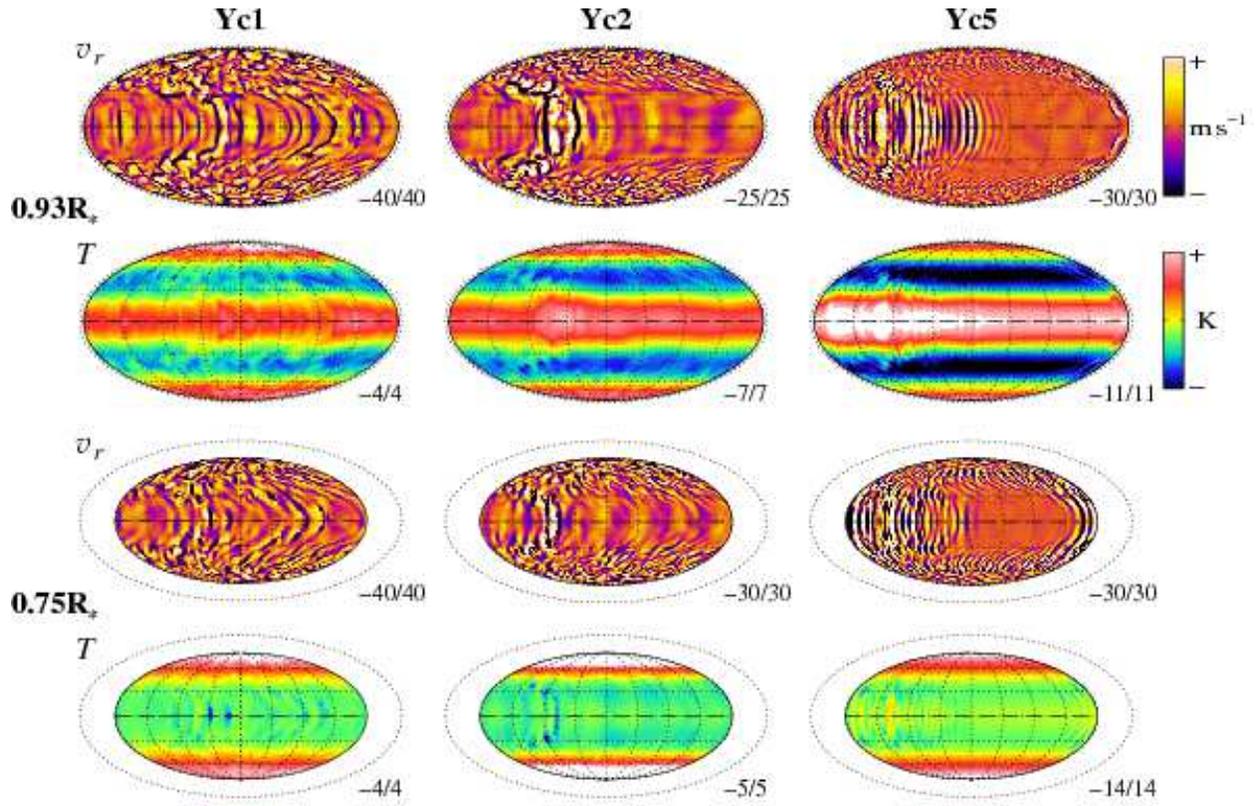}
\figcaption{Fields $v_r$ and $T$ near the surface and at mid-depth, for the three models Yc1, Yc2, and Yc5. See caption of Fig.~\ref{Fig:shsl_Ya}.\label{Fig:shsl_Yc}}
\end{figure*}

\begin{figure}
\epsscale{1.0}
\plotone{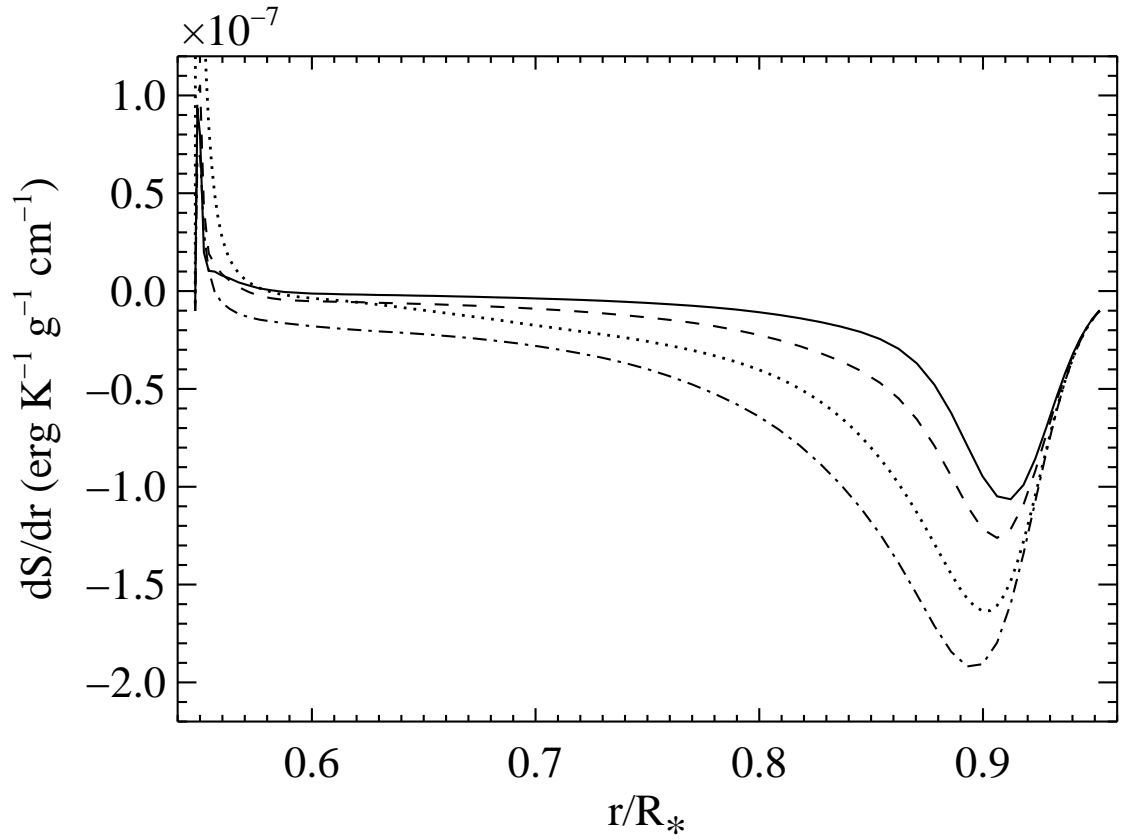}
\figcaption{Entropy gradient $d\Sbb /dr$ for models Ya1 ({\itshape solid line}), Ya2 ({\itshape dashed line}), Ya5 ({\itshape dash-dotted line}), and Ya5T ({\itshape dotted line}).\label{Fig:dsdr_Ya}}
\end{figure}

We now compare the Ya series with the series of simulations Yc to see the effects of decreasing the Prandtl number from
$P_r=4$ to $P_r=1/4$.
Convective patterns of the three main Yc simulations are shown Fig.~\ref{Fig:shsl_Yc}, analog of Fig.~\ref{Fig:shsl_Ya}.
In going from the Ya1 to the Yc1 case, the Taylor number is increased, as it was when $\Omega_o$ was increased. However, in this case there is no visible stabilization effect since, thanks to the reduction of $\nu$, the turbulence level increases as well. Yc1 is the most turbulent run that we performed, if we exclude the special cases Yc5T and Yd5, as shown by the Reynolds numbers $R_e$ (Table~\ref{Tab:descrip}). Banana cells are less visible and
structures are more complicated. 
Relative to Ya1, the asymmetry of up/down flows is more pronounced in Yc1.
The small-scale vorticity is more intense in this model than in more laminar ones, as suggested by a higher Rossby number $R_o$ (Table~\ref{Tab:descrip}). By comparing, on $v_r$ maps, the convective patterns in the polar and equatorial regions, it becomes clear
in this simulation that the behaviors of convection in these regions are different. We can roughly separate what occurs inside and outside the cylinder tangent to the inner shell and parallel to the rotation axis. Inside this cylinder, convection follows a ``polar regime'' slightly influenced by rotation, outside it follows an ``equatorial regime'' where rotation influences the convection much more.

When the rotation rate $\Omega_o$ is increased, convective features are modified.
We can see  that in the equatorial region convection tends to be localized, favoring a limited band of longitudes at a given time.
By contrast, at the solar rotation rate, this is not very pronounced. We note that around the equator there is a zone (close to the map center) where the velocity is slightly stronger. When the rotation is double or more (Yc2, Yc5), vigorous convective motions are concentrated in a small region in longitude. In the rest of the equatorial
zone, convective velocities are markedly smaller (by a factor 5 or 10 typically).
Polar convection is much less affected. We observe that, in this region, convective patterns are smaller, for reasons we have previously developed. Thus, convective motions inside and outside the vertical tangent cylinder are only loosely connected. 

Finally, we can compare once again $v_r$ and $T$ maps in Fig.~\ref{Fig:shsl_Yc}. Correlations are apparently less obvious. This is because the axisymmetric component of $T$ presents a stronger contrast than for Ya1 or Ya5 (it is quite similar to Ya5T). If the $m=0$ components are removed from $T$, correlations clearly appear again.
We retrieve in these model runs the same structure of azimuthal bands as in the previous series. This is a  characteristic common to all our model simulations. Only the thermal contrast of these bands varies. It as been increased here by a factor 4$\sim$5 relative to Ya series. As for the previous series, the thermal contrast increases with $\Omega_o$.

\subsection{Pattern temporal evolution}

\begin{figure}
\epsscale{1.0}
\plotone{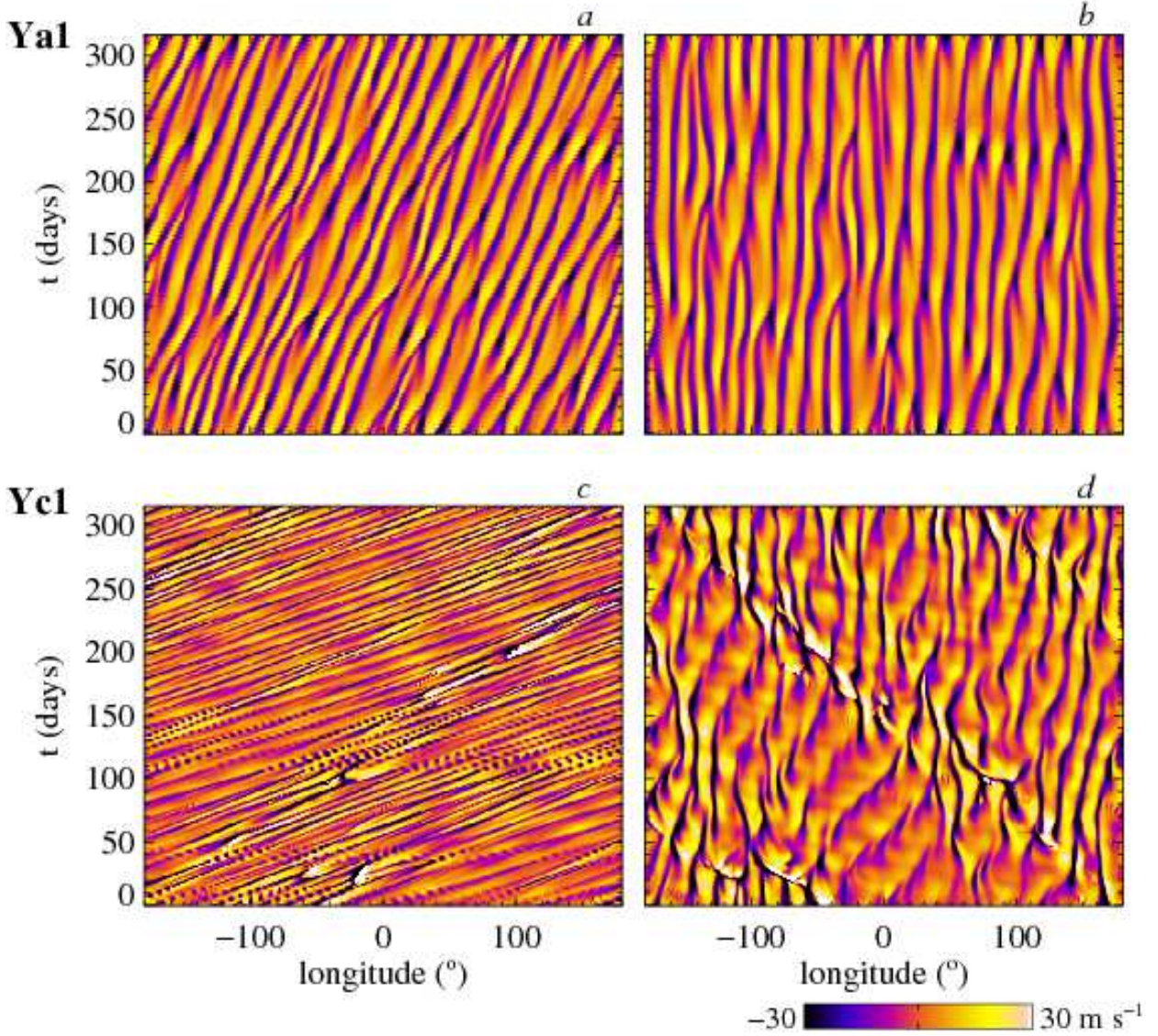}
\figcaption{$(\phi,t)$-diagrams showing the temporal evolution of the $v_r$ profile along the equator ($\theta_c=90\degr$), close to the top of the shell ($r_c=0.93\:R_*$) for models Ya1 ($a,b$) and Yc1 ($c,d$). On the left ($a,c$), diagrams are plotted in the corotating frame. On the right ($b,d$), the diagrams are plotted in the ``shifted'' $(\phi_s,t)$ coordinates, to remove the local effect of differential rotation (see text).\label{Fig:pattern}}
\end{figure}

After having described the spatial structures of convection in our simulations, we discuss here how they evolve with time.
To do so, we have followed the evolution of convective patterns thanks to a time-longitude ($\phi$,$t$) diagram. Figure~\ref{Fig:pattern} shows such diagrams for Ya1 and Yc1 models.
For given radius $r_c$ and zenith angle $\theta_c$, we have plotted the radial velocity field $v_r$ in the $(\phi,t)$ plane, serving to follow, for each longitude, the temporal evolution of $v_r$.
We have also plotted ``shifted" diagrams, in $(\phi_s,t)$ coordinates, with
\begin{equation}
\phi_s=\phi-\frac{1}{r_c}\int_{t_0}^t \Vph (r_c,\theta_c,t) \mathrm{d}t
\end{equation}
where $\Vph$ is the azimuthal average of $v_\phi$, in order to compensate for the local advection of the differential rotation.

A first look at the non-shifted maps (Figs~\ref{Fig:pattern}$a$ and \ref{Fig:pattern}$c$) shows that velocity structures propagate eastwards. It is still moderate for Ya1, but becomes really strong for Yc1, so much so that the map becomes difficult to read. For such cases it is very useful to consider shifted maps since they are more readable.
Since the maps are plotted in the corotating frame, such an eastward propagation is a signature of a strong differential rotation, where the equator is accelerated relative to the mean rotation speed
(cf. \S~\ref{sec:DR}).

Let us focus on Ya1. Maps show convective patterns are persistent  during several turnover times: we can follow them in the map during the complete time sequence. Figure~\ref{Fig:pattern}$b$ shows that differential rotation does not account for the total eastward propagation. There is still a small residual drift. We thus see a self-propagation of convection structures. For example, the structure located at $\phi=-80\degr$ at the initial time is around $-60\degr$ longitude 300 days later.
We clearly see also the regular appearance of downflow lanes and their ongoing mergers.
These maps show undoubtedly the temporal modulation of speed and thickness of flows, as a consequence of the intrinsic fluctuation of convective motions.

It is instructive to make a direct comparison with the pattern evolution seen in Yc1 simulation. The shifted map of this turbulent model (Fig.~\ref{Fig:pattern}$d$) reveals
conspicuous temporal modulations, stronger than in Ya1. Moreover, the persistence of downflow lanes is shorter.
The hot spots, described in previous section, are persistent patterns. On the shifted map, they obviously drift westward, a signature of retrograde motion.

The localized convective structure seen on Yc2 velocity maps (Fig.~\ref{Fig:shsl_Yc}) is also persistent and presents the same kind of retrograde motion (not shown).
In the Yc5 simulations, the localized convective patterns also similarly propagate, but their temporal evolution is very peculiar: the convective level fluctuates highly in this case, so much so the pattern periodically fully disappears, reappearing later. Section~\ref{Sec:OscConv} is dedicated to the study of this special case.

\section{Differential rotation produced}\label{sec:DR}
We have seen in the previous section that convective patterns propagate in longitude under the influence of a large-scale mean longitudinal flow. Here, we discuss in more detail this mean flow: the differential rotation, and how it varies, both in contrast and profile, with the rotation rate.

\subsection{Differential rotation profiles}\label{ssec:DRprof}

\begin{figure*}
\epsscale{.85}
\plotone{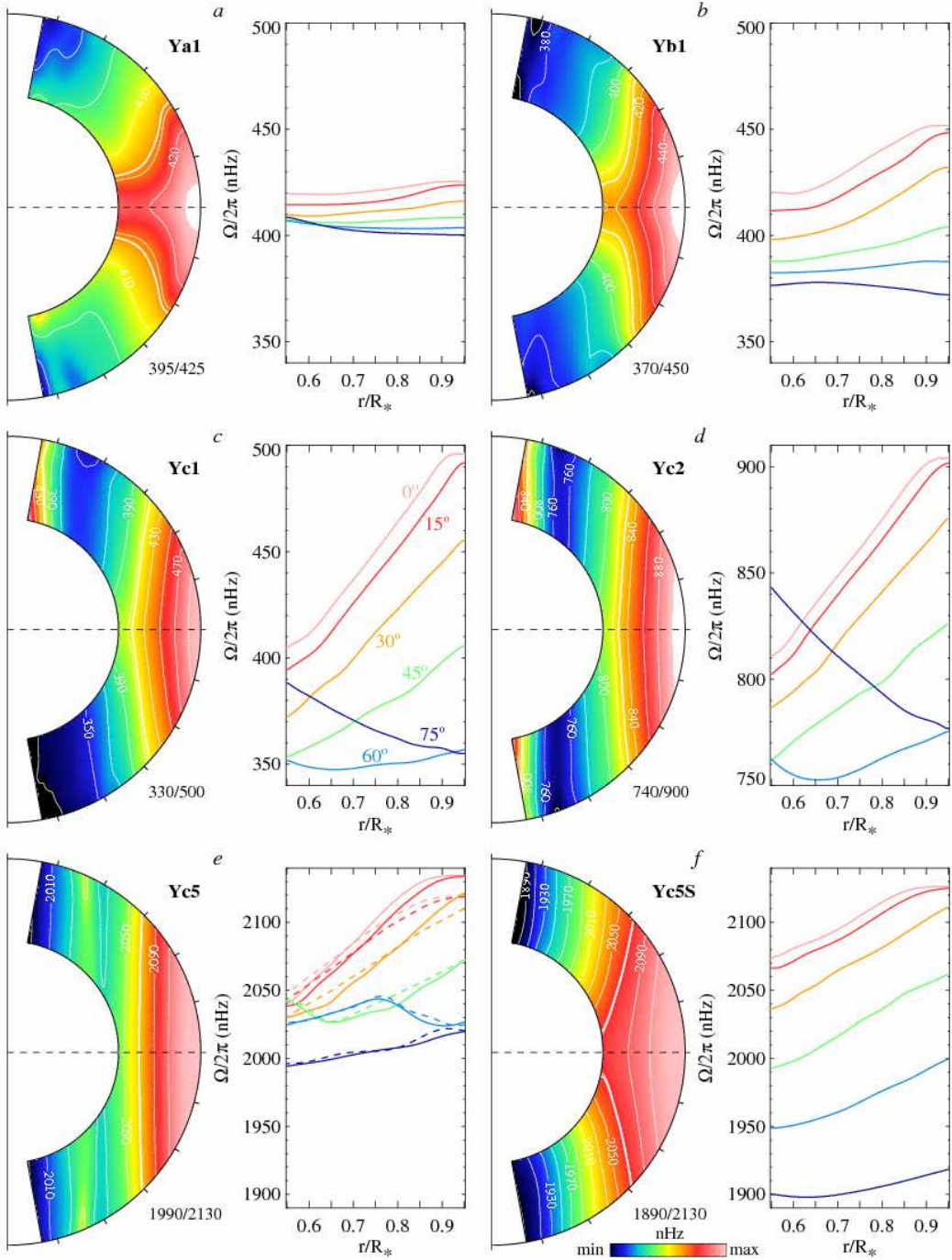}
\figcaption{Temporal and azimuthal averages of angular velocities achieved for six simulations: Ya1, Yb1, Yc1, Yc2, Yc5, and Yc5S. For each panel $a$--$f$, a contour plot of $\Omega/2\pi$ in the meridional plane is shown on the left, and on the right, radial cuts of the same quantity are plotted at six fixed latitudes (specified on panel $c$), averaged on both hemispheres. The color scale of each map is independent and is indicated nearby the plot. Thick lines on contour plots correspond to $\Omega=\Omega_o$. The scales of the right plots are the same for panels $a$--$d$, and also for panels $e$ and $f$. Since Yc5 is an oscillating model, the $e$ right panel shows radial cuts of $\Omega/2\pi$ at two different moments: when the differential rotation reaches 1) a maximum ({\it solid lines}) and 2) a minimum ({\it dashed lines}).
\label{Fig:omY}}
\end{figure*}

Figure~\ref{Fig:omY} shows the temporal and longitudinal averages of angular velocities achieved for six representative models. To make comparison with observations easier, we have converted mean azimuthal velocities $\Vph$ into sidereal angular velocities $\Omega=\Omega_o+\Vph/(r\sin\theta)$. 
The rotation profile of our reference case, Ya1 (Fig.~\ref{Fig:omY}$a$), exhibits a clear equatorial acceleration, similar to the Sun. All of our models present such a behavior. It is a consequence of the domination of the Coriolis force over the buoyancy \citep{Gilman77}, as measured by the low convective Rossby number : Table~\ref{Tab:descrip} indicates that $R_{oc}<1$ in every case.
In Ya1, the contrast of this differential rotation is not very large, but the decrease of $\Omega$ from the equator to the poles is monotonic (see right panel). The rotation profile is rather conical, i.e., radial cuts at different latitudes are independent of $r$.

On Fig.~\ref{Fig:omY}, we can see the effects of increasing $T_a$, following two different paths : first by decreasing the viscosity $\nu$ and so $P_r$ : simulations Ya1 to Yc1 via Yb1 (panels $a$, $b$, $c$); second by increasing $\Omega_o$ in going from Yc1 to Yc5 via Yc2 (panels $c$, $d$, $e$).

In going from Ya1 to Yb1, we first notice the increase of the contrast of the differential rotation (Fig.~\ref{Fig:omY}$b$), multiplied roughly by a factor of 4. The profile remains monotonic with respect to the latitude coordinate. However, it becomes less conical, and tends to becomes more cylindrical, i.e. with isorotation contours parallel to the stellar axis. Compared to Yb1, in Yc1 (Fig.~\ref{Fig:omY}$c$), the profile becomes even more cylindrical and the contrast is increased by a factor of 2. Such a contrast of 130$\:$nHz is typical of those observed at stellar surfaces (around 90$\:$nHz for the Sun).
The rotation rate decreases monotonically from equator up to 60\degr, but the monotonicity is lost in the polar caps. Moreover the behavior of the rotation is different in the northern and southern polar caps. As we have already seen in \S~\ref{sec:conv} concerning the convection, dynamics inside and outside the tangent cylinder are practically disconnected, and the two polar regions are disconnected too. However, very long averages should result to identical profiles.

By following the path \#2 (Yc1, Yc2, Yc5), the cylindrical aspect is more and more marked and the isorotation contours for Yc5 model (Fig.~\ref{Fig:omY}$e$) are almost purely vertical. The rotation behavior in the polar regions varies from one model simulation to the next in sequence, from one hemisphere to the other. Nevertheless, outside the tangent cylinder, the equatorward monotonic increase of $\Omega$
is always present.
The contrast of the differential rotation is slightly reduced along this path.
Even though it is not very pronounced for this Yc series, it is more noticeable for Ya and Yb model series. Thus, models with high $P_r$ and high $\Omega$ reach a rather weak differential rotation. The extreme case, Ya5, rotates practically as a solid body.

Along both paths, when $T_a$ is increased, rotation profiles become more cylindrical.
However, the contrast of the differential rotation responds differently in the two cases: it rises along path \#1 and declines when $\Omega_o$ increases. This is an effect of the stabilizing role of rotation, an effect already mentioned in the previous section. This means that a non negligible part of the properties of linear unstable modes is retained in these non-linear simulations.

We will discussed of the rotation profile of Yc5S case and the temporal modulation of Yc5 profile further in \S~\ref{Sec:OscConv}.

\subsection{Rotation scaling law}

\begin{figure}
\epsscale{1.0}
\plotone{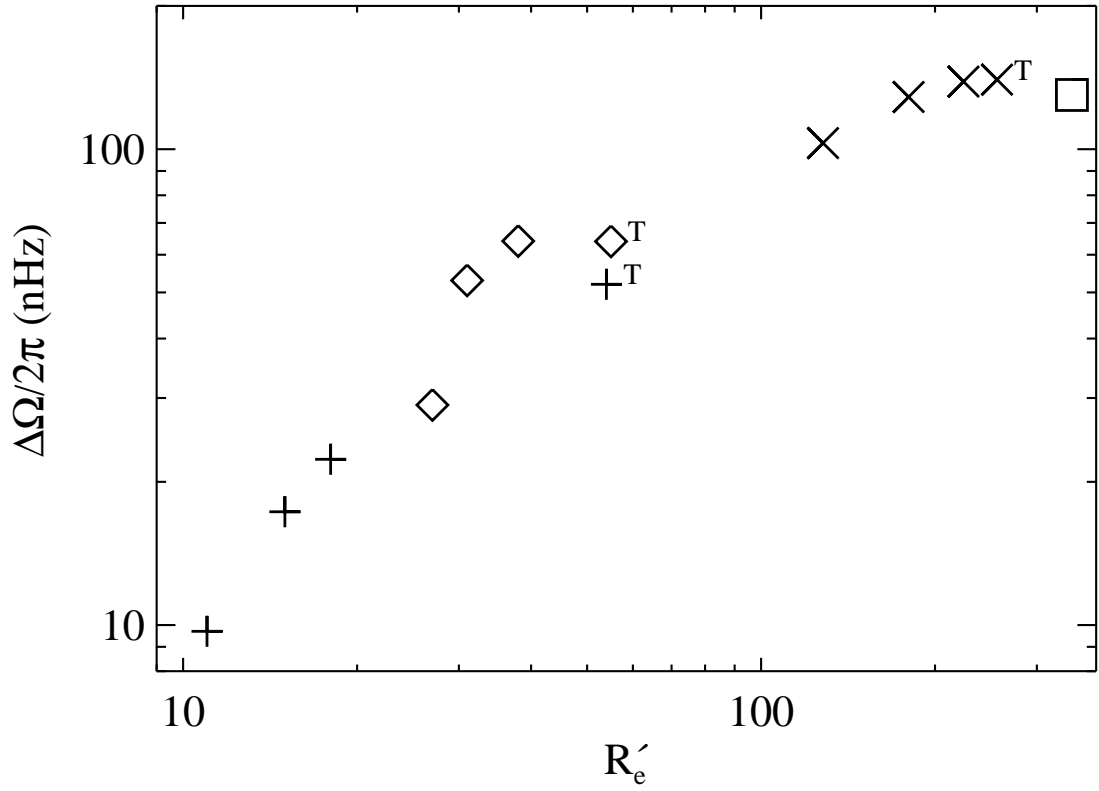}
\figcaption{Differential rotation contrasts $\Delta\Omega$ according to convective Reynolds numbers $R_e'$.
Plus, diamonds, crosses, and square correspond to Ya, Yb, Yc, and Yd models respectively.
T indexes indicate models Ya5T, Yb5T, and Yc5T.
\label{Fig:DomRe}}
\end{figure}

\begin{figure}
\epsscale{1.0}
\plotone{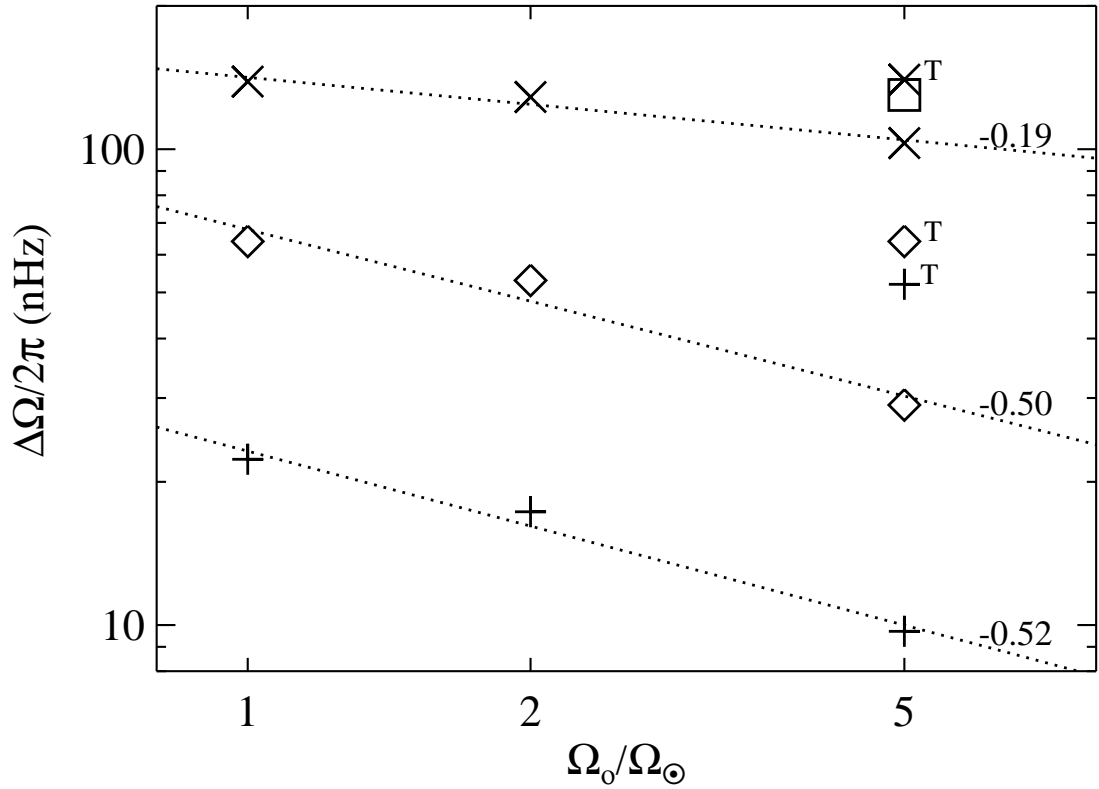}
\figcaption{Differential rotation contrasts $\Delta\Omega$ according to rotation rates $\Omega_o$.
The sense of symbols is the same as in Fig.~\ref{Fig:DomRe}.
Dotted lines correspond to the scaling law deduced by logarithmic regression for each series. Values of fitted slopes are printed out on the right.\label{Fig:DomOm}}
\end{figure}

We have seen that differential rotation contrasts can vary with variations in the model parameters.
To quantify these variations we have computed for each model the contrast $\Delta\Omega$ at the surface of the convective shell, between the equator and the latitude 60\degr (Figs.~\ref{Fig:DomRe} \&\ref{Fig:DomOm}).
We have chosen to compute it near the surface because That is where rotation is observed in stars (cf. \S~\ref{Sec:intro}). Moreover, by choosing to cut the range at 60\degr\ we avoid the polar regions where peculiar behavior occur in Yc series. These contrasts $\Delta\Omega$ are summarized in Table~\ref{Tab:veloc} among other useful quantities. We have plotted $\Delta\Omega$ as a function of the turbulence level characterized by Reynolds number $R_e'$ (Fig.~\ref{Fig:DomRe}), and of the rotation rate $\Omega_o$ (Fig.~\ref{Fig:DomOm}) for all our models. Thus we get a global vision of the sensitivity of  $\Delta\Omega$ to different parameters.

First, there is a clear and strong dependence of $\Delta\Omega$ on the turbulence level. The simulation Ya5T is more turbulent than Ya5 --- with the same $P_r$ and $\Omega_o$ --- and reaches a stronger contrast, even higher than Ya1 (see Table~\ref{Tab:veloc}). That is also true when we compare Yb5T to Yb5 or Yc5T to Yc5.
To show the strong impact of turbulence level, we can compare $\Delta\Omega$ and $R_e'$. 
We have defined two Reynolds numbers: $R_e=\Vvr L/\nu$ and $R_e'=\Vvr' L/\nu$ (see Table~\ref{Tab:descrip}). The former is computed with the rms velocity at mid-depth layer $\Vvr$; the latter is computed with $\Vvr'$, which is the rms velocity of convective motions, i.e. that evaluated after removing the azimuthal averages to the velocity components (Table~\ref{Tab:veloc}). It is more relevant to compare $\Delta\Omega$ with $R_e'$ than with $R_e$: when the differential rotation is strong, $\Vph$ is the major contribution to $\Vvr$ and so increases the value of $R_e$, thus, by construction, we have to find a correlation between $\Delta\Omega$ and $R_e$. On the contrary,  $R_e'$ takes into account the convective motions only, since the differential rotation has been filtered for its evaluation. Thus by comparing $\Delta\Omega$ to $R_e'$, we really measure the impact of the convective turbulence level on the differential rotation.
Figure~\ref{Fig:DomRe} shows the clear direct correlation between $\Delta\Omega$ and $R_e'$ we have found. The contrast $\Delta\Omega$ increases with $R_e'$ and seems to saturate around 140$\:$nHz in the more turbulent cases. The signification of this correlation will be discussed in \S~\ref{ssec:amom}.

The variation of $\Delta\Omega$ by going from Ya to Yc, seen in Table~\ref{Tab:veloc}, seems to indicate a dependence to $P_r$. However, this variation is mainly the effect of the growth of the turbulence level from Ya to Yc. Thus we have to compare models with same turbulence levels, such as Ya5T and Yb5T. The latter, with a lower $P_r$, achieves a slightly higher differential rotation than the former. Nevertheless, by comparing Yc5T to Yd5, one could conclude the opposite. In any case, the sensitivity to $P_r$ is obviously weak in comparison with $R_e'$.

Finally, we have carefully studied the effects of $\Omega_o$ on $\Delta\Omega$.
For each series Ya, Yb and Yc, we have deduced a scaling law from every set of three simulations having the same parameters except $\Omega_o$ (for instance the set Ya1, Ya2, \& Ya5). We have fitted a power law
$\Delta\Omega\propto\Omega_o^\alpha$. The resulting fits are plotted in Fig.~\ref{Fig:DomOm}. The $\alpha$ exponent increases 
from around $-0.5$ for the more laminar simulations (Ya and Yb) to $-0.19$ for the more turbulent ones (Yc).
Thus for every series $\Delta\Omega$ decreases with the rotation rate, more or less strongly.
When $\Omega_o$ is increased, due to stabilizing effects $R_e'$ decreases and therefore $\Delta\Omega$ too.
This phenomenon could explain the sign of the slope. However, the fall of $R_e'$ is not the only effect: by comparing Yb1 and Yb5T, we notice that, at $5\,\Omega_\sun$, the turbulence level must be higher ($R_e'\sim55$ for Yb5T against 38 for Yb1) to achieve the same contrast ($\Delta\Omega\approx64\unit{nHz}$).
If we focus now on the Yc series, the decrease in this case is less pronounced. All of the models achieve contrasts between 100 and 140\unit{nHz}, which is close to observations. Turbulence seems to be sufficiently developed to make the contrasts of this set of models less sensitive to $R_e'$. The exponent of the scaling law we have obtained with this more reliable sequence is only slightly negative ($-0.19$).
Despite differences in details, this is not far from the most recent results of \citet{BarnesC05} cited in the introduction. They have found a marginally positive coefficient which is mainly compatible with the relation $\Delta\Omega\sim{}$constant. If we extrapolate our sequence Ya, Yb, Yc to more turbulent series with lower $P_r$, we expect that $\Delta\Omega$ tends to be independent of $\Omega_o$.

\section{Meridional circulation}\label{sec:MC}

The meridional circulation is markedly smaller than the differential rotation, but can play a non-negligible role in the dynamics, especially in the heat and angular momentum transport (see \S~\ref{ssec:amom}). Further, its topology is an important parameter for models of stellar dynamos, such as the Babcock-Leighton type ones \citep[e.g.,][]{JouveB07}.

\begin{figure*}
\epsscale{1.0}
\plotone{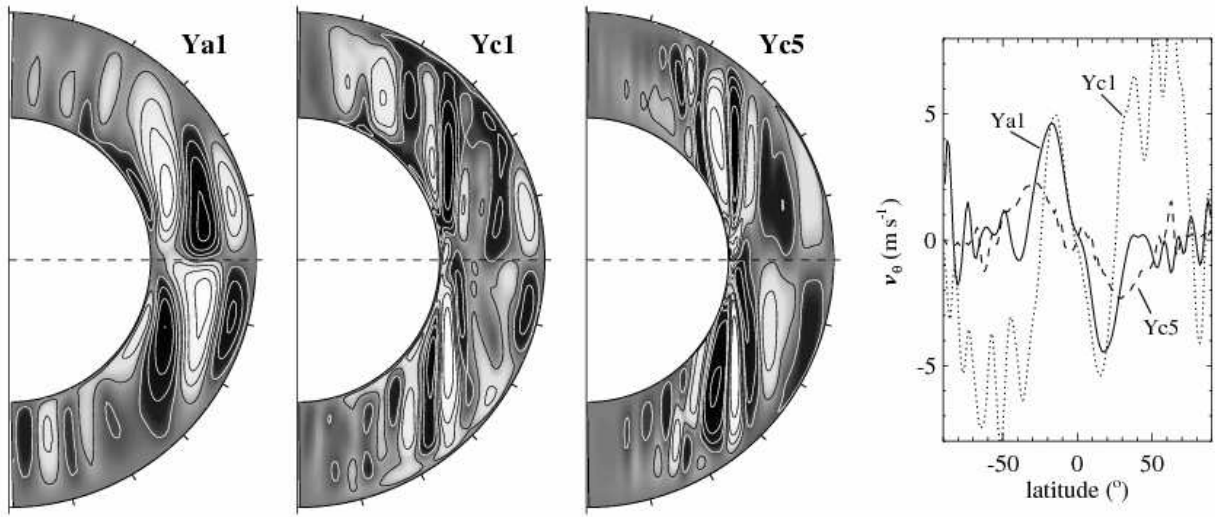}
\figcaption{Meridional circulation in Ya1, Yc1, and Yc5. Maps show the mean mass flux circulation in the azimuthal plane. White lines over black background correspond to clockwise circulations, and black lines over white background to counterclockwise ones. The color scale is not linear to make visible even weak flows. Mean latitudinal velocity $\Vth$ profiles at the top of the shell are also plotted. Maps and profiles are obtained by averaging over longitude and time.\label{Fig:CM}}
\end{figure*}

Figure~\ref{Fig:CM} shows the time-averaged meridional circulations produced in three of our model simulations: Ya1, Yc1 and Yc5. We have plotted in the meridional plane the mass flux streamfunction, as defined in \citet[][eq.~{[7]}]{MieschE00}. We have also plotted on this figure the velocity of the circulation at the top of the domain. Since our boundary is impenetrable, the radial component $\Vrh$ vanishes, so $\Vth$ is the total speed of the meridional circulation. The amplitude of the flow is a few\unit{m\,s^{-1}}, ranging between 1 and 10 for the set of simulations considered in this study. Due to the compressibility effects, deep inside the convecting shell, velocities decrease strongly to conserve the mass flux in denser regions.

Considering first the reference case Ya1, we see a well-organized multicellular circulation, where
both hemispheres present almost antisymmetric cells. If we exclude what happens around the poles, there are four cells along the radius. The cells near the top of domain induce poleward flows at the surface near the equator, that are similar to the solar observations and simulations.
Near the poles, cells are smaller, vertically oriented, and go across the entire thickness of the modeled shell.

Turning now to Yc1 model with a higher $T_a$, we notice a more intricate and complex topology with an increase of the number of cells. The cells near the surface in the equatorial region retain the same circulation direction and a similar intensity, even if they are a little less extended.
We notice that meridional circulation becomes stronger along the tangent cylinder. In this region, the structures become more ``concentrated'' and long and thin cells appear.
If we continue to increase $T_a$ by increasing $\Omega_o$, this effect is reinforced; long cells near the tangent cylinder are more concentrated and dominate in intensity over others further away from the tangent cylinder. We can find some similarities between this phenomenon and the Stewartson layers seen in simulations of rotating stellar radiative regions \citep{Rieutord06}.
In the equatorial region, the cells already seen both for Ya1 and Yc1, are still present, having greater extent in latitude, but weaker velocity.

\begin{figure}
\epsscale{1.0}
\plotone{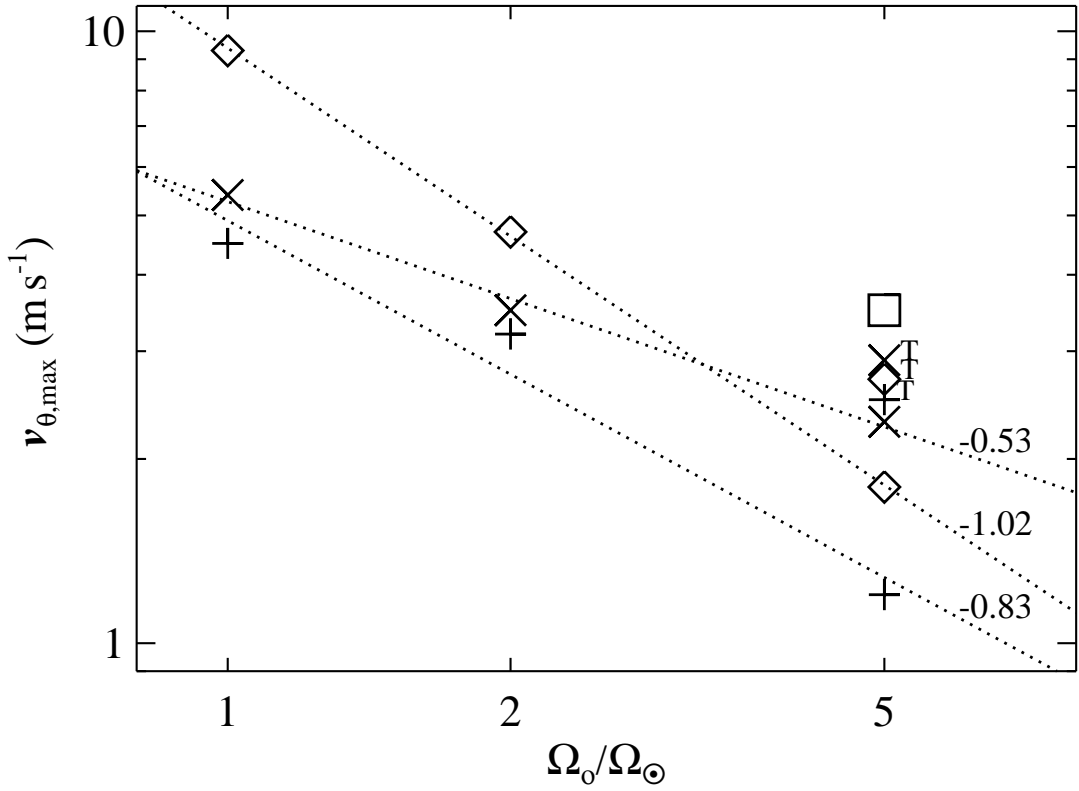}
\figcaption{Surface velocity $\Vth{}_\mathrm{max}$ characterizing the meridional circulation intensity, plotted as function of the rotation rate $\Omega_o$. See caption of Fig.~\ref{Fig:DomOm}.\label{Fig:mclaw}}
\end{figure}

To show the dependence of the meridional circulation on the rotation rate, we have scale it by the maximum velocity of the equatorial cells, present in all models. We plot the maximum of $\Vth$ at the surface in these cells as a function of $\Omega_o$ in Fig.~\ref{Fig:mclaw}.
This velocity is generally the highest $\Vth$ produced at the surface of our models, except for Yc1 (but that does not change our conclusion).
We see that $\Vth{}_\mathrm{max}$ in the Yb series are higher than in both Ya and Yc for rotations 1 and 2 times the solar rotation rate. This hierarchy between series is different from the one observed for $\Delta\Omega$. However, if we should have considered the total kinetic energy in the meridional circulation (see MCKE in Table~\ref{Tab:veloc}), the hierarchy between Yb and Yc should be the same as for the differential rotation. The difference has been mentioned before: in Yc series, a large part of MCKE is concentrated deep inside, along the tangent cylinder, and is not visible in $\Vth{}_\mathrm{max}$.

As previously done with $\Delta\Omega$, power laws are fitted to the three sets of model simulations with identical input parameters, excepting $\Omega_o$. For all series, the power-law exponent is clearly negative (between $-0.5$ and $-1$).
In contrast to what we have seen for differential rotation, this trend of declining $\Vth{}_\mathrm{max}$ with $\Omega_o$ is rather strong. Even if in Yc5T, the more turbulent counterpart of Yc5, the intensity of meridional circulation slightly increases relative to Yc5, but it stays at a lower level than both Yc1 and Yc2, a property that was not found for $\Delta\Omega$.
Thus, if the differential rotation decreases only marginally for high rotation speeds, the meridional circulation clearly weakens according to our simulations.

\section{Interpreting the dynamics}\label{sec:interp}

Our simplified model of a rapidly rotating convective envelope yields a 
nonlinear dynamical system with complex feedbacks. 
Numerical simulations are a very efficient tool to assess the dynamical balances
that are achieved in such systems. They provide detailed insights in the way 
kinetic energy, heat and angular momentum are exchanged among the different reservoirs.
Given the importance of understanding the
internal differential rotation profile of stars, it is necessary to 
identify the processes involved in establishing these large scale flows.
In the following we discuss how the differential rotation and meridional
circulation discussed in \S~\ref{sec:DR} and \S~\ref{sec:MC} are produced.

\subsection{Energy balance}\label{ssec:energy}

We start by analyzing the energy balance of the various simulations discussed in the paper.
In Table~\ref{Tab:veloc} we have summarized key quantities such as rms velocity or volume averaged kinetic
energy (KE${}=\frac{1}{2}\rb(v_r^2+\Vt^2+\Vp^2)$) achieved in the models. In order to have a more precise analysis, we decompose the rms
velocities into their three components, with or without subtracting the azimuthal averages, and KE into its axisymmetric and
non-axisymmetric components, as in \citet{BrunT02}. We define the kinetic energy in differential rotation 
as DRKE${}=\frac{1}{2}\rb\Vph^2$,
the kinetic energy in the meridional circulation as MCKE${}=\frac{1}{2}\rb(\Vrh^2+\Vth^2)$,
and the kinetic energy in non-axisymmetric (convective) motions as CKE${}={}$KE${}-{}$DRKE${}-{}$MCKE.

Turning first to the rms velocities we notice a large range of values, from a few up to about one hundred \unit{m\,s^{-1}}, especially for $\Vvr$ and $\Vpr$. Both these velocities have the same magnitude in every model;
in contrast, when comparing the longitudinal velocity $\Vpr$ to its radial and latitudinal counterparts, we see significant differences: $\Vpr$ can be up to 10 times higher than $\Vrr$ and $\Vtr$. This is mostly due to the large scale differential rotation achieved in almost all the simulations
and most certainly in the Yc and Y\#T series as seen in \S~\ref{sec:DR}. If we instead compare the rms velocities with
their axisymmetric mean subtracted, we find that the three components have about the same amplitude with 
$\Vrr' \sim \Vrr$ being the fastest. This yields convective flows that are largely isotropic. If we define
the isotropy index as $a_r=\Vrr'{}^2/\Vvr'{}^2$,
we find that it varies from about $1/2$ in the Ya series to about $1/3$ in the Yc series. At large Prandtl
number the anisotropy of the flow is thus greater than when $P_r$ is small. This can perhaps be explained
by the fact that the flows are more turbulent in the latter series, having both larger Reynolds and Rayleigh numbers
(see Table~\ref{Tab:descrip}). This leads to convection that is less dominated by large scale rolls (consistent with the
disappearance of banana cell structures) that are characterized by large coherent radial motions. Instead,  the convection, characterized by more
chaotic random motions, is dominated by strong downward convective plumes as discussed in \S~\ref{sec:conv}. Within one of these
plumes downwards vortical motion dominates, but when forming horizontal averages their small filling factor do not
contribute as much as large scale coherent rolls. By increasing $\Omega_o$, $a_r$ tends to rise slightly, probably for similar reasons. The stabilizing effects of rotation also reduce the supercriticality of flows. In this case, convection is less vigorous and more influenced by large scale rolls.

Studying the partition of kinetic energy between its axisymmetric and non-axisymmetric components is also very instructive.
As was shown in Fig.~\ref{Fig:evolFen}, our simulations, after a short linear convective instability phase, undergo a nonlinear 
saturation and finally reach, in most of the cases after about 2000~days, a statistically stationary state,
over which meaningful temporal and volume averages of the kinetic energy and its mean and fluctuating contributions
can be evaluated. Turning again to Table~\ref{Tab:veloc}, we notice that the energy in the meridional flows is 
very weak (less than 0.2\% of KE) compared to DRKE and
CKE that contain most of the energy. Again, this is because the meridional flows in such systems
result from small imbalances between several large terms and are less easily developed than azimuthal motions \citep{Miesch05}.
Another obvious trend, present in all the series Ya, Yb and Yc is that an increase of the rotation 
rate leads to a smaller kinetic energy available for the system. This reduction is about a factor of
3 for Ya and Yb series and only 1.6 for the Yc series. Rotation thus stabilizes
convection and leads to weaker mean or fluctuating flows, as the decrease in absolute value of DRKE, CKE and MCKE 
indicates (note however that in terms of percentages CKE \& DRKE remains about the same). 
The fact that the simulation in the Yc series are less sensitive to the increase of $\Omega_o$ is certainly due
to their higher degree of supercriticality. We also find that the ratio of fluctuating and mean (mostly 
azimuthal) motions changes in favors of the latter as we decrease the Prandtl number. In the Ya series with
$P_r=4$, DRKE represents only about 35\% of KE, with most of the energy concentrated in CKE. By contrast
DRKE represents respectively about 80\% and 95\% for Yb and Yc series. It is thus clear that the higher is 
the Taylor number, the stronger is DRKE and the associated differential rotation, as the column $\Delta \Omega_\mathrm{top}$
also clearly indicates (see also discussion in \S~\ref{sec:DR}). Given the strong stabilizing effect of $\Omega_o$ on the motions, we have also performed simulations with rotation set at 5 times the solar rate, but with a higher degree of surpercriticality (models labeled with T).
All these models posses a large DRKE, including model Ya5T. Thus the fact that DRKE is found to increase with decreasing
$P_r$ is perhaps more linked to the supercriticality of the models than to the ratio between 
viscous and thermal diffusivities. A useful number to assess the influence of rotation on convection
is the convective Rossby number $R_{oc}$ (see Table~\ref{Tab:descrip}). This number does not include $\nu$ or $\kappa$. We find that this number is always relatively small in the simulations 
with the largest differential rotation contrast, indicating a dominant effect of rotation over convection,
meaning that as the convective flows reverse direction they are significantly tilted by the Coriolis force. 

In the Yc series, for the models rotating at 5 times the solar rate, the convection develops a rather intermittent behavior,
in both space and time. Computing time-averaged quantities such as KE and rms velocities can thus be misleading, since they depend strongly on the state
of the convection (quiescent or excited). In cases like Yc5 and Yc5T, we find that at the peak of the convection
burst CKE can account for almost 30\% of KE before weakening to 5\%, whence the differential rotation
has developed again. We discuss these simulations in more details in \S~\ref{Sec:OscConv}.

\subsection{Angular momentum balance}\label{ssec:amom}

The differential rotation discussed in \S~\ref{sec:DR} results from angular momentum redistribution
within the shell of intense convection realized in our simulations. 
In purely hydrodynamical models the differential rotation profile is the result of a competition among viscous torques, Reynolds stresses and meridional circulation.
By adopting stress-free boundary conditions at the top and bottom boundaries of
our simulations, no net torque is applied to these rotating spherical shells of
convection.  Thus total angular momentum within our simulations is
conserved, and we can examine the manner in which it is redistributed, following the
approach developed in \citet{ElliottM00} and \citet{BrunT02}.

\begin{figure}
\epsscale{0.45}
\plotone{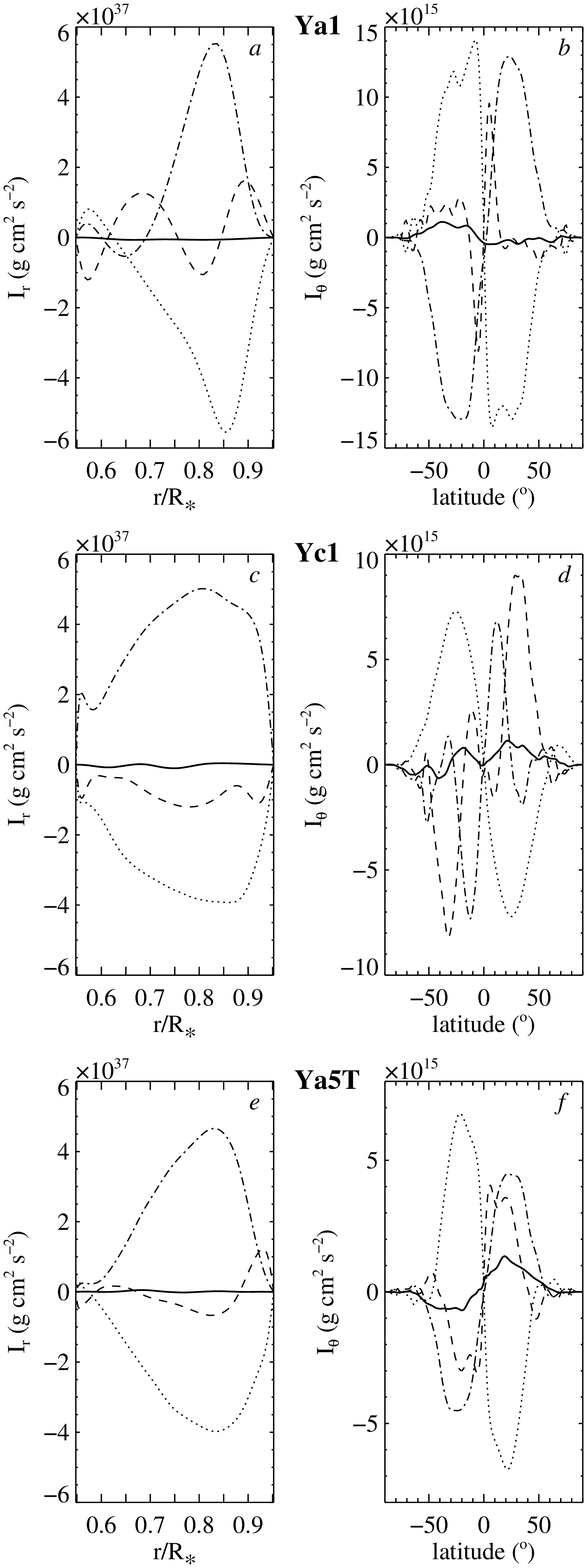}
\figcaption{Time average of the latitudinal integral of the radial angular momentum flux $I_r$ ({\itshape left}), and  of the radial integral of the latitudinal angular momentum flux $I_\theta$ ({\itshape right}), for models Ya1 ({\itshape top}), Yc1 ({\itshape middle}) and Ya5T ({\itshape bottom}). The fluxes have been decomposed into their viscous ({\itshape dotted lines}), Reynolds stress ({\itshape dot-dashed lines}), and meridional circulation components ({\itshape dot-dashed lines}). The total fluxes are plotted with solid lines.\label{Fig:amom}}
\end{figure}

The temporal derivative of angular momentum averaged over longitude can be written as a divergence:
\begin{equation}
\frac{\partial {\cal L}}{\partial t}
= \frac{1}{r^2} \frac{\partial (r^2 {\cal F}_r)}{\partial r}
+ \frac{1}{r \sin\theta} \frac{\partial (\sin\theta {\cal F}_\theta)}{\partial \theta}.
\end{equation}
with ${\cal F}_r$ and ${\cal F}_\theta$ being respectively the radial and latitudinal components of 
the averaged angular momentum flux.
They can be written and separated into their three contributions
(due again to Viscous torque, Reynolds stresses and Meridional Circulation):
\begin{equation}
\begin{array}{lcr}
{\cal F}_\theta = {\cal F}_{\theta,V} + {\cal F}_{\theta,R} + {\cal F}_{\theta,MC}&&
\\
& = \hat\rho r \sin\theta \Big[
-\nu \frac{\sin\theta}{r} \frac{\partial}{\partial\theta}
\left( \frac{\Vph}{\sin\theta}  \right)
+ \widehat{v'_\phi v'_\theta}&\\
&&+ \Vth (\Vth + \Omega_o r \sin \theta )\Big] 
\end{array}
\end{equation}
\begin{equation}
\begin{array}{lr}
{\cal F}_r= {\cal F}_{r,V} + {\cal F}_{r,R} + {\cal F}_{r,MC}&\\
&=\hat\rho r\sin\theta \Big[
-\nu r \frac{\partial }{\partial r}\left(\frac{\Vph}{r}\right)
+\widehat{v'_\phi v'_r}
+\Vrh \left(\Vph+\Omega_o r \sin\theta\right)
\Big]
\end{array}
\end{equation}
We can integrate the $\theta$-flux along concentric spheres of varying radius and the $r$-flux along cones
with various latitudinal inclinations as follows:
\begin{equation}
I_{\theta,X}=\int_{r_b}^{r_t} {\cal F}_{\theta,X} r \sin \theta \mathrm{d}r \qquad
X=V,R,MC.
\end{equation}
\begin{equation}
I_{r,X}=\int_{0}^{\pi} {\cal F}_{r,X} r^2 \sin \theta \mathrm{d}\theta \qquad
X=V,R,MC.
\end{equation}
Since most of the simulations are statistically stationary, the temporal average
$\langle\partial {\cal L}/{\partial t}\rangle_t$ vanishes. In Figure~\ref{Fig:amom}
we display the time-averaged integrated fluxes $I_{r}$ and $I_{\theta}$ and their components for 
the models Ya1, Yc1 and Ya5T, in order to assess the effect of varying $P_r$ or/and $\Omega_o$
in the overall angular momentum balance.

Turning first to the radial fluxes of angular momentum (Figs.~\ref{Fig:amom}$a$, $c$ and $e$),
we see that viscous forces act to transport angular momentum radially
inward in all cases. 
This inward transport is opposed mainly by the Reynolds stress flux. The role of the meridional circulation 
is less systematic. In cases Ya1 and Ya5T the MC flux is alternatively positive and negative
following with good fidelity the multicellular profiles maintained in these cases and shown in Fig.~\ref{Fig:CM}.
In Yc1, the meridional circulation contribution is actually the strongest of all the simulations, with
a characteristic amplitude two to three times larger than in most of the other cases.  This simulation possesses a rather
large and dominant clockwise cell in most of the domain which transports angular momentum
inward at low latitude. This results in a meridional circulation flux that helps
the viscous stresses to slow down the surface and speed up the base of the convection zone,
thus opposing the Reynolds stresses, to yield a total radial
angular momentum flux that is nearly zero, as noted in Fig.~\ref{Fig:amom} by the solid
thick lines.  While the systems here are highly variable in time, by allowing the
system to evolve for extended periods of time (typically thousands of days)
and performing long time averages, we appear to be sensing the equilibrated
states reasonably well. In going from the mildly turbulent flows of case Yc1 or Ya5T to
the complex ones of case Yc1, we see that the viscous flux has dropped and
that the Reynolds stresses and meridional circulations have changed
accordingly to maintain equilibrium.  The Reynolds stresses in cases Yc1 and Ya5T
are more evenly distributed over the whole depth of the domain, whereas in the more laminar
case Ya1 is is mostly concentrated near the surface.

Examining now the latitudinal transport of angular momentum (Figs.~\ref{Fig:amom}$b$, $d$ and $f$),
we see that the effect of the Reynolds stresses in both cases is primarily
to speed up the equator, since ${\cal F}_{\theta,R}$ is positive in the
northern hemisphere and negative in the southern.  It is opposed by the viscous fluxes, 
which act to speed up the poles and to enforce solid body rotation. 
The MC flux helps the Reynolds stresses by speeding up the equatorial region.
This effect is opposite to what is found in simulations of the present day 
Sun with its thinner shell of convection \citep{BrunT02}.
In case Ya1 the MC flux is localized near the equator while in the two other
cases these fluxes spread over a larger latitudinal band. 
In particular the circulation in model Yc1 possesses a strong
equatorward cell from low latitude ($\sim 20\degr$) up to about $70\degr$, 
which dominates the poleward cell near the surface.
This larger contribution of the MC flux in this model, confines the Reynolds
stresses toward lower latitudes.
The manner in which each of the different components of ${\cal
F}_{\theta}$ acts does not appear to vary appreciably in going from one case
to another.  There are thus no clear trends in following either path \#1 (lower $P_r$) 
or path \#2 (higher $\Omega_o$; as described in \S~\ref{ssec:DRprof}).
However as the level of complexity is increased, we see a
decrease in the magnitude of all components of ${\cal F}_{\theta}$.

We conclude that for these cases, involving either a low $P_r$ or a 
strong rotational constraint, the Reynolds stresses act latitudinally to speed 
up the equator, and radially to slow down the base of the convection zone by transporting
angular momentum from the bottom of the domain toward the top.
The latitudinal Reynolds stresses are opposed by viscous effects, whereas they
are aided somewhat by the meridional circulations.  On the contrary the radial
Reynolds stresses have to act against both viscous and meridional flux transport.

The overriding role of Reynolds stresses to achieve an equatorial acceleration explains, at least in part, the correlation between $\Delta\Omega$ and $R_e'$ discussed in \S~\ref{sec:DR} and shown Fig.~\ref{Fig:DomRe}.
Here, $R_e'$ can be seen as a measurement of the balance between viscous and Reynolds stresses effects.
When $R_e'$ rises, Reynolds stresses are more efficient in transporting the angular momentum, since the flow is less viscous.

This detailed balance gives us a picture of how the angular momentum is continuously
transported, and which processes dominates acting to speed up or
slow down certain regions.  However, it is difficult to infer from the
fluxes alone what the actual $\Omega$ profile (either cylindrical or conical as in the Sun) 
will be, besides stating that the equator will be fast or slow from the sense of the viscous fluxes
(which are always down the gradient of $\Omega$).

\subsection{Thermal wind}\label{ssec:thw}
It is well known that fluids under the influence of strong rotation can become quasi 2-D
with most of their dynamics invariant with respect to dimension parallel to the rotation axis 
\citep{Pedlosky}. Indeed if Coriolis forces dominate the others, it can be shown that:
\begin{equation}\label{eq:TPT}
\frac{\partial \Vph}{\partial z} = 0
\end{equation}
which corresponds to a barotropic configuration. It is a consequence of the 
so-called Taylor-Proudman theorem.
However, rotating convection involves both radial and latitudinal heat transport,
with the likelihood that latitudinal gradients in temperature and entropy
may result within the convective zone.  This implies that surfaces of mean
pressure and density will not coincide, thereby yielding baroclinic terms
in the vorticity equations \citep{Pedlosky,Zahn92}.  Under sufficiently
strong rotational constraints, a `thermal wind balance' might be achieved
in which departures of the angular velocity from being constant on
cylinders (aligned with the rotation axis) are controlled by those
baroclinic terms.  Indeed, some mean-field approaches have invoked such a
balance to obtain differential rotation profiles with bearing on the solar
convection zone \citep[e.g.,][]{KitchatinovR95, Rempel05a}. As discussed in \citet{BrunT02} and \citet{MieschB06} in more details, such a balance effectively implies that
\begin{equation}\label{eq:tw2}
\frac{\partial\Vph}{\partial z} =
\frac{1}{2\Omega_o\hat{\rho}c_{P}}
\nab\hat{S}\times\nab\hat{P}\Bigr|_{\phi}
= \frac{g}{2 \Omega_o r c_{P}} \frac{\partial\hat{S}}{\partial \theta}
\mbox{ ,}
\end{equation}
where $z$ is parallel to the rotation axis.  Thus latitudinal entropy
gradients could serve to break the Taylor-Proudman constraint which would
otherwise require the rotation to be constant on cylinders \citep{Rempel05a,MieschB06}. 
However it is important to note that this constraint may also be broken by Reynolds and viscous 
stresses.  
Indeed the strong correlation found between $\Delta\Omega$ and $R_e'$ (cf. Fig.~\ref{Fig:DomRe}) is a clear indication that Reynolds stresses are key players for the redistribution of angular momentum.
To strengthen this point, we show below that these terms are as important as the baroclinic terms in establishing the
differential rotation in the bulk of the convective envelope.

\begin{figure*}
\epsscale{1.0}
\plotone{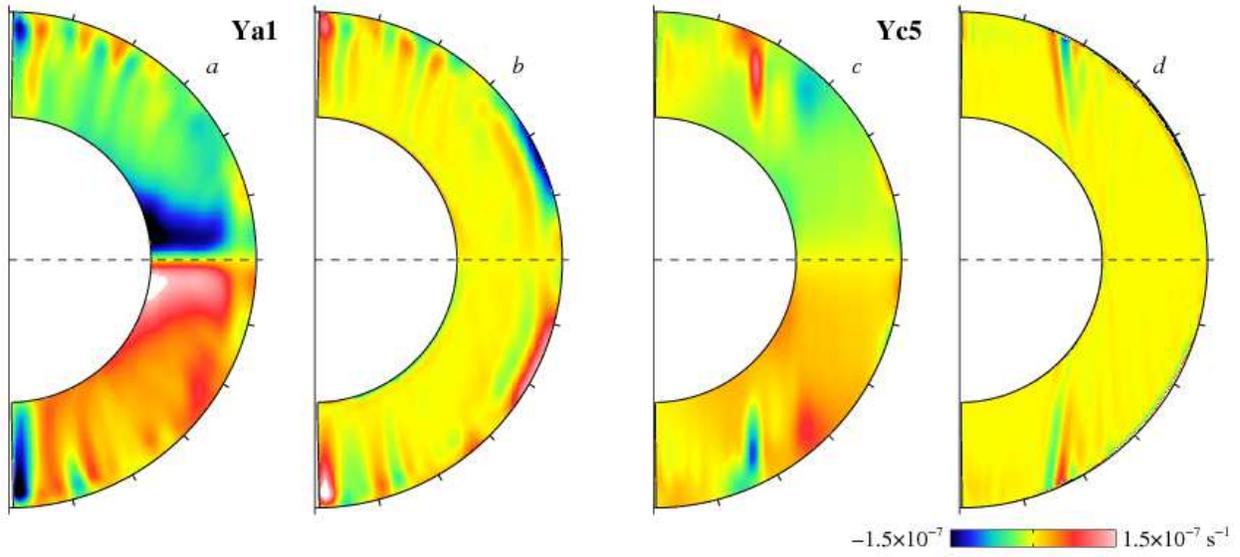}
\figcaption{Temporal and azimuthal averages for models Ya1 ($a,b$) and Yc5 ($c,d$) showing the baroclinic term in the meridional force balance ---defined eq.~(\ref{eq:tw2})--- ($a,c$), and the difference of this term with $\partial \Vph/\partial z$ ($b,d$).\label{Fig:thw}}
\end{figure*}

Figure~\ref{Fig:thw} assesses for case Ya1 and Yc5 the extent to which latitudinal entropy
gradients serve to drive the temporal mean zonal flows $\Vph$
seen as the differential rotation in Fig.~\ref{Fig:omY}. Figures~\ref{Fig:thw}$a$ and \ref{Fig:thw}$c$ display the latitudinal entropy 
term seen on the right hand side of equation (\ref{eq:tw2}), which in an exact thermal wind 
balance would be identical to $\partial\Vph/\partial z$.
Figures~\ref{Fig:thw}$c$ and \ref{Fig:thw}$d$ show the difference between this baroclinic term and the
actual $\partial\Vph/\partial z$, thus providing a measure of
departures from such a balance.  Within most of the entire convection zone thermal wind
dominates in both cases, meaning that there is a close relationship between the zonal thermal
wind and the differential rotation established in our simulations. 
However, near the top of the domain we do see large departures from thermal wind balance.
Further we note that for the rapidly rotating case Yc5, in the region near the tangent cylinder
a strong shear layer (also clearly visible in Figure~\ref{Fig:CM} showing the meridional circulation)
is present which is not in thermal wind balance. There viscous and Reynolds stresses play
crucial roles, not unlike what is seen in simulations of rotating stellar radiation zones,
the so-called Stewartson layers \citep[e.g.,][]{Rieutord06}.

Thus examining the nature of this thermal wind balance, combined with the
assessment of fluxes of angular momentum, reveals that Reynolds, meridional flows and viscous
stresses have a major role in establishing the differential rotation in the 
convection zone, with help coming from latitudinal thermal gradients.
As seen in Figures~\ref{Fig:shsl_Ya} \& \ref{Fig:shsl_Yc} the thermal wind is associated with weak latitudinal temperature 
variations.  In Table~\ref{Tab:veloc} we quote the contrast of $T$ at the base of the unstable domain.
We find that $\Delta T_\mathrm{bot}$ varies from 2$\:$K in the most laminar cases up to 15$\:$K in the most turbulent
ones. We also see that increasing $\Omega_o$ leads to larger temperature contrasts.
However, as it can be seen in equation (\ref{eq:tw2}), in order to retain a strong baroclinic term and
thus a differential rotation profile that is not constant along cylinders, $\Delta S$ (and 
the associated $\Delta T$) must increase by exactly the same factor. But, as
we can see in Table~\ref{Tab:veloc}, this is hardly the case, with for instance case Yc5 having
a contrast a factor of 2 to 3 larger than Ya5 instead of a factor of 5 that would exactly compensate the increase of the rotation rate. Thus the profiles become 
more cylindrical ($\partial\Vph/\partial z \rightarrow 0$) even though the flow remains 
somewhat in thermal wind balance. 
Thus, if rapidly rotating stars do not rotate along cylinders, they need an efficient latitudinal heat transport that establishes a
strong entropy (temperature) contrast. Indeed, to get conical solar-like rotation 
profiles,
it is not sufficient to have an established and balanced thermal wind (which should 
nevertheless develop after several dynamical times), but it is necessary to have a strong baroclinicity in the model. Such large gradients could be favored by the presence of a shear layer
at the base of the convective domain, such as the solar tachocline \citep{MieschB06}. 
We will indeed show in \S~\ref{ssec:thforcing} that enforcing a large entropy variation does lead to less cylindrical 
rotation profiles.

\section{Oscillating convection}\label{Sec:OscConv}

In this section, we focus on the oscillating behavior of convection, found in some of our simulations with rapid rotation,  that we have mentioned several times through this paper. Models with low Prandtl numbers and high rotation rates, like Yc5, are typical of these vacillating convective solutions. We present here in more detail this phenomenon and discuss the physical processes governing it.

\subsection{From localized to vacillating convection}
We have described in \S~\ref{sec:conv} that equatorial convection tends to be localized for Yc models. Another phenomenon occurs in case Yc5: temporal modulations become really huge and the convection vacillates. The polar regions are not affected, convection there is uniformly developed without temporal variation.  In the polar caps, the rms convective velocity $\Vvr'$, in midlayer depth, remains  around 24\unit{m\,s^{-1}}. However, in the equatorial area, convection, as well as being localized, oscillates between a state of fully-developed convection and a quiet state, characterized by very slow motions during which the convection is ``asleep''. In the equatorial band $\Vvr'$ fluctuates between 3 to 45\unit{m\,s^{-1}}.

These oscillations are clearly not a transitional state before a statistically stationary state, as those discussed in \S~\ref{ssec:1stevol} and shown Fig.~\ref{Fig:evolFen}. The Yc5 simulation is relaxed, but oscillates. This kind of behavior has already been observed in Boussinesq simulations of the geodynamo \citep{GroteB01}, and is observed here for the first time in simulations of deep stellar convection \citep[see also][]{BrownB07}.

\begin{figure}
\epsscale{1.0}
\plotone{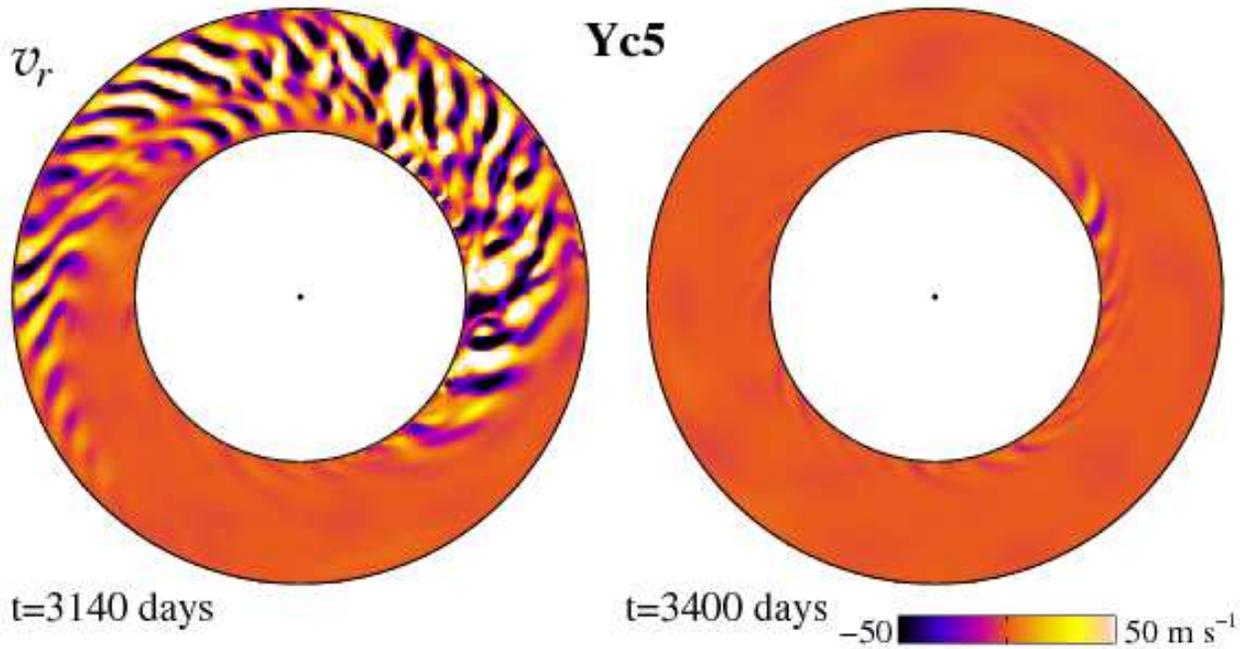}
\caption{Radial velocity field in the equatorial plane for model Yc5 during a maximum ({\itshape left}) and a minimum ({\itshape right}) of convective activity.\label{fig:eqYc5}}
\end{figure}

\begin{figure}
\epsscale{1.0}
\plotone{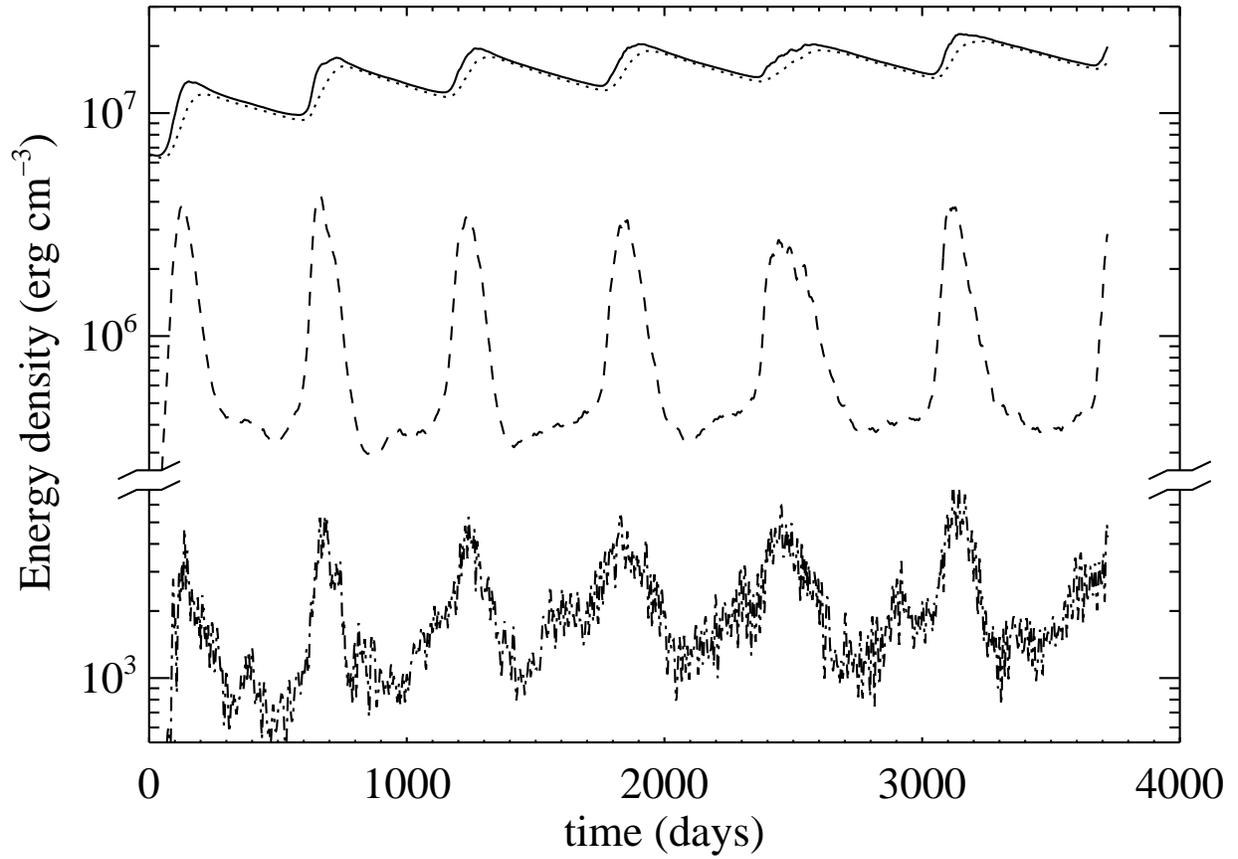}
\figcaption{Temporal evolution of kinetic energy content for the Yc5 case. KE ({\itshape solid line}) is decomposed into its components: DRKE ({\itshape dotted line}), CKE ({\itshape dashed line}), and MCKE ({\itshape dot-dashed line}).\label{Fig:evolYc5}}
\end{figure}

Figure~\ref{fig:eqYc5} shows convective patterns at two different epochs in the equatorial plane of the star, illustrating clearly the bimodal (high-low) state of convection. During convective bursts (Fig.~\ref{fig:eqYc5} \textit{left}), convection is well developed in the equatorial plane, even though it is developed only on one half of the area, similar to what we see in Yc2 case.
During quiet periods (Fig.~\ref{fig:eqYc5} \textit{right}), convection is practically fully stopped everywhere around the equator.
We can follow these oscillations in the kinetic energy (KE) evolution (cf. Fig.~\ref{Fig:evolYc5}). 
This variability is almost periodic and the period is around 600~days. The averaged enthalpy flux (Eq.~[\ref{eq:Fen}], not plotted here) shows also oscillations very similar to CKE.
Convective bursts last around 150~days, and CKE is quite flat between each burst.
The variation of differential rotation is different. DRKE varies like ``sawteeth". It begins to increase quickly during the burst, so the growth lasts for 
$\sim 150$ days. Then this phase is followed by a slow decaying period of $\sim 450$ days, before a new cycle starts. As a consequence, we clearly see a phase difference between
the maximums of DRKE and CKE.

Left panel of Fig.~\ref{Fig:omY}$e$ allows us to see how the DRKE variation is distributed in latitude.
We notice that, for latitudes greater than 50\degr, $\Omega$ is approximately uniform and does not change with time. Cuts of $\Omega$ at latitudes 45, 60 and 75\degr\ do not show sensible differences between a maximum and a minimum of DRKE. This is consistent with the fact that in these polar regions convection is practically not modulated.
Variations occur mainly around the equatorial band in the upper layers. It induces variations in $\Delta\Omega$ of almost 15$\:$nHz. The main change occurs in the slope of $\Omega$ profile along the radius, mainly at latitude 0\degr\ and 15\degr. Such variations imply also variations of the radial shear in the equatorial area.

To investigate further, we have performed the more turbulent models Yc5T and Yd5. These model runs need higher resolution and thus are very expensive in terms of computing time. Yc5T is a model with the same Prandtl number as Yc5 but with a higher Reynolds number.
It presents the same behavior with a slightly longer period ($\sim 680$ days): DRKE-decrease phase lasts for $\sim 450$ days, as in case Yc5, and its bursts, slightly longer, going on for about 230 days typically.
The persistence of the oscillating phenomenon in this more turbulent simulations indicates that it is not due to marginally stable conditions. All of these simulations are clearly supercritical.
Another extra run, Yd5, as turbulent as Yc5T but with $P_r=1/8$ shows once again oscillations. However, the period reaches $\sim 1060$ days, the burst phase length is roughly the same as for Yc5T case, but decrease duration is of $\sim 800$ days.
$P_r$ seems to influence inversely the decrease time, whereas turbulent level seems to govern the burst duration and so DRKE-increase length.
We also notice that simulations with higher $P_r$, like Ya5 or Ya5T, do not show vacillating convection. This is because these cases have lower $T_a$ (see Table~\ref{Tab:descrip}). The Taylor number appears to be a key parameter governing the oscillations; they are triggered for all simulations with a Taylor number above a threshold of $T_a \sim 10^9$. Below this threshold, no models oscillate.

Using all these results, we can build a sketch of the oscillating phenomenon.
When the convection develops, the Reynolds stresses $\langle v_i'v_j' \rangle$ 
become stronger, driving a larger differential rotation.
In the equatorial zone, the radial shear increases (Fig.~\ref{Fig:omY}$e$).
The shear is so strong that it destroys the coherence of convective cells (visible at midlayer depth  in Fig.~\ref{fig:eqYc5}), killing the convection.
Thus $\langle v_i'v_j'\rangle$ terms practically vanish; this implies that the
differential rotation decreases by viscous dissipation.
When the shear is sufficiently low in the equatorial region, convection can amplify again,
and a new cycle begins.

During quiet phases, the part of energy which is not evacuated by convection is piled into internal energy. During bursts, the kinetic energy is pumped from the reservoir of internal energy, which is on several orders of magnitude larger.
With magnetic effects, we can imagine very complicated situations. Oscillation properties can 
drasticly change, even disappear \citep{GroteB01,MorinD04}.

\begin{figure}
\epsscale{1.0}
\plotone{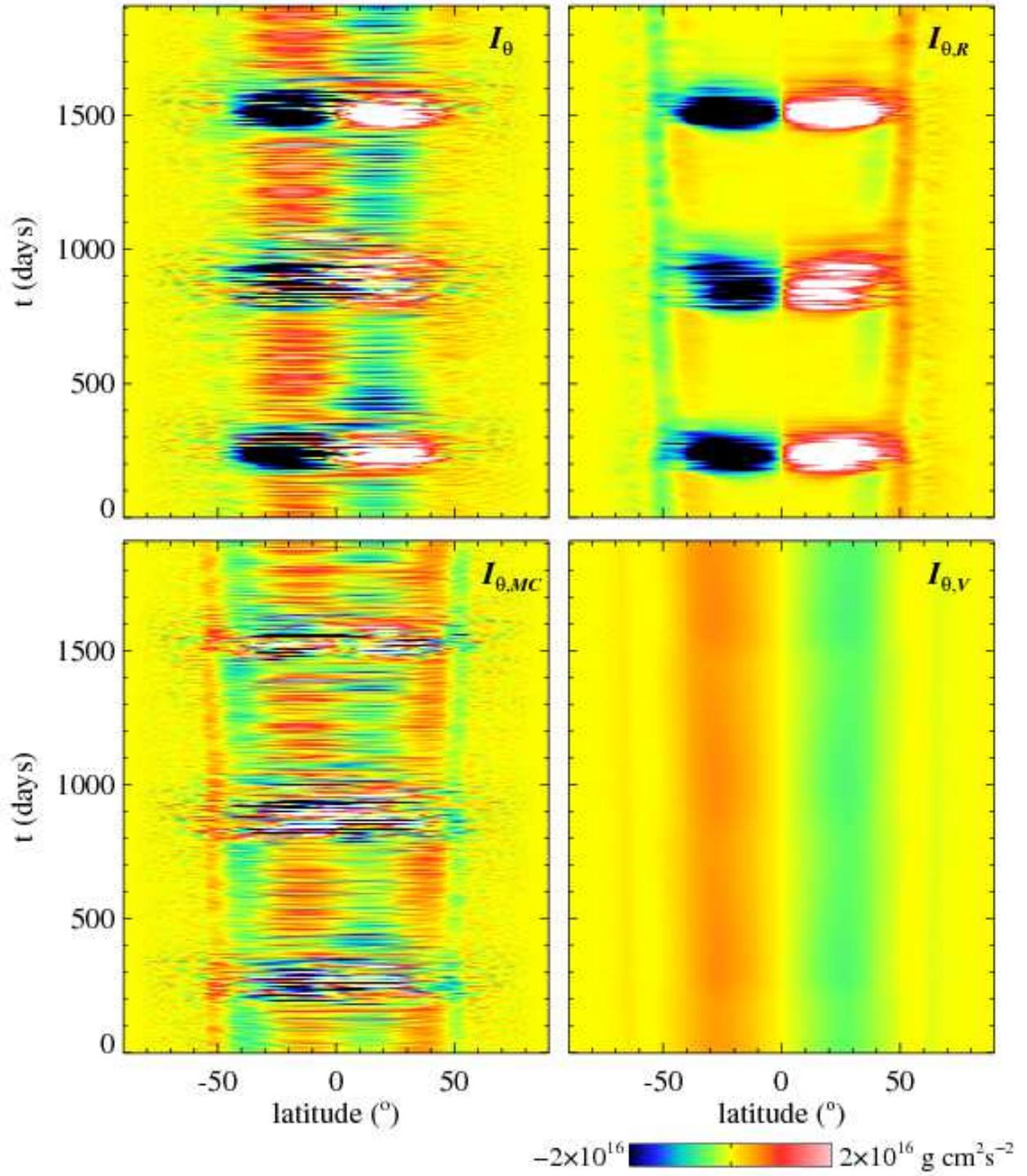}
\caption{Evolution as a function of time of $I_\theta$, the angular-momentum flux in $\theta$-direction, averaged on radius.  $I_\theta$ ({\itshape top left}) is decomposed into the contributions of Reynolds stresses $I_{\theta,R}$ ({\itshape top right}), meridional circulation $I_{\theta,MC}$ ({\itshape bottom left}), and viscous torque $I_{\theta,V}$ ({\itshape bottom right}). White/red (black/green) indicate southward (northward) transfer.\label{fig:FluxCinYc5}}
\end{figure}

A way to verify this scenario is to follow, during the evolution, the fluxes of angular momentum defined in \S~\ref{ssec:amom}.
Figure~\ref{fig:FluxCinYc5} shows the evolution of each integrated flux $I_{\theta}$ along few oscillation cycles. We can see that near the poles ---i.e., inside the tangent cylinder--- momentum flux and all of its components are roughly null. We clearly see as expected that during bursts (there are three of them on the plots) the contribution of Reynolds stresses dominates and drags angular momentum toward the equator. Viscous effects, rather constant with time, act to decelerate the equator when convection has damped out, in agreement with the scenario described above. Effects of the meridional circulation are harder to assess because it fluctuates more, especially during the bursts where the flux is rather large but its sign unceasingly changes. During quiet states, it helps the viscous dissipation to decelerate the equator, but compensates viscous effects at middle latitudes. This is a consequence of the multicellular topology of the meridional circulation (cf. Fig.~\ref{Fig:CM}).

\subsection{Influence of boundary conditions: thermal forcing}\label{ssec:thforcing}
Rotation profiles of these rapidly rotating simulations are very cylindrical, like those predicted by the first solar simulations
before they were negated by helioseismic observations. Up to now, there has been no observational inference of the inner rotation profile for other stars. Thus, such cylindrical profiles could reflect reality, but we cannot exclude the possibility that our simulations suffer problems similar to those of the first solar simulations.
In this section, we want to see how the dynamics is modified when the ``cylindricity'' of rotation profiles is reduced. Indeed, if the rotation profiles of thermally forced models are more conical than those of previous unforced models, the shear in the equatorial plane would have to be reduced, along with the oscillation process.
Profiles of the Yc5 model are almost cylindrical, due to the high rotation rate.
Coriolis forces dominate the others and the equation~(\ref{eq:TPT}) is almost verified.
The Taylor-Proudman theorem can be broken if a baroclinic term exists, as defined by the equation (\ref{eq:tw2}). However, as seen in \S~\ref{ssec:thw} this term has reduced in strength in going from Yc1 to Yc5.
Thus a potential thermal wind can lead to more conical profiles (see discussion \S~\ref{ssec:thw}). The origin of such a thermal
wind could be a strongly sheared region, like the tachocline observed in the Sun \citep{Rempel05a,MieschB06}.

We have imposed such an entropy gradient at the base of the convective shell in a twin of Yc5 simulation: Yc5S. The latitudinal entropy profile imposed at the bottom of the shell is in the form
\begin{equation}
\frac{S_\mathrm{bot}}{c_P}=a_2Y_2^0+a_4Y_4^0,
\end{equation}
with $a_2=1.14\times 10^{-5} $ and $a_4=-1.8\times 10^{-5}$.
In the present configuration, compared to solar case \citep{MieschB06}, profiles are strongly cylindrical and the rotation speed is five times higher. So we need to impose a strong gradient to expect a sensible effect. In this simulation the contrast of temperature between poles and equator is around 60$\:$K, that is 5 times larger than the Sun, 9 times larger than Yc1 case, and 3.5 larger than the initial Yc5 model (see also Table~\ref{Tab:veloc}).

\begin{figure}
\epsscale{1.0}
\plotone{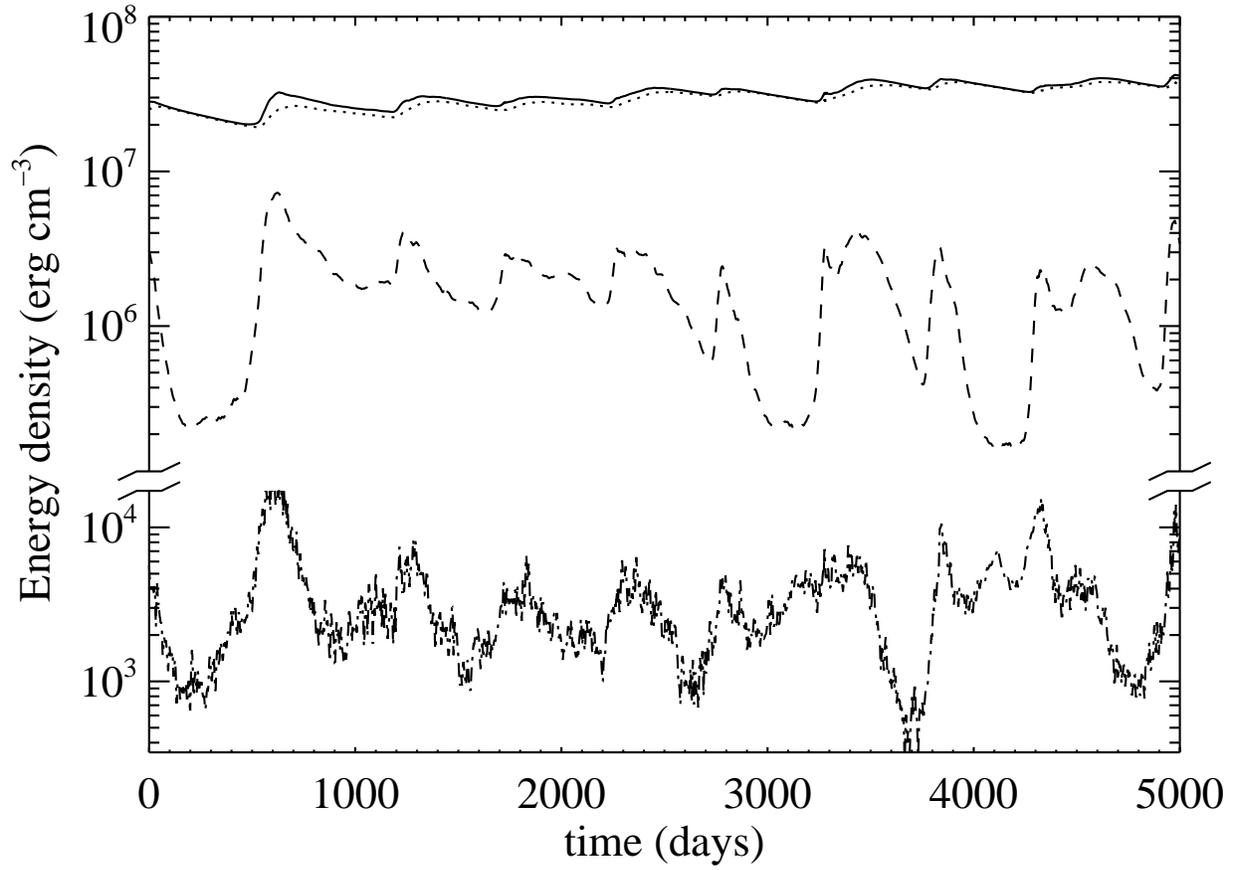}
\figcaption{Temporal evolution of kinetic energy content for the Yc5S case. See caption of Fig.~\ref{Fig:evolYc5}.\label{Fig:evolYc5S}}
\end{figure}

Figure~\ref{Fig:evolYc5S} shows the temporal evolution
of kinetic energy.
After a cycle similar to those shown by Yc5, the entropy has diffused and we reached a phase 
characterized by a reduction of the oscillation amplitudes, as we had expected.
Figure~\ref{Fig:omY}$f$ shows this profile at $t\sim 2000$~days. The rotation is less cylindrical, even less than Yc1 case. As a consequence, the contrast of differential rotation
along radius decreases in the equatorial zone, which is clearly seen by comparing the slopes of curves at $0\degr$ in the right panels of Fig.~\ref{Fig:omY}$e$ and \ref{Fig:omY}$f$.
As a function of latitude, the contrast of rotation increases and the profile becomes purely monotonic: the decrease of rotation continues at high latitudes ---i.e., inside the tangent cylinder--- and does not stop around $50\degr$ as in Yc5.
However, after $t\sim 2600$~days, our model achieves a rather chaotic new state (Fig.~\ref{Fig:evolYc5S}). There are strong variations in the convective state again, but no more simple periodicity. Indeed, differential rotation has continued to increase in latitude, but in radius also, making possible a sufficiently strong shear again.

As expected, imposing an entropy gradient at the base of the convective zone could lead to a more conical rotation profile, modifying the dynamics and reducing the oscillations.
The thermal contrast needed can be caused by a strong shear due to a thinner tachocline as was shown by \citet[][eq. {[11]}]{Brun07}.
Indeed, in order to get a more conical profile for the same mean radial jump in angular velocity in the tachocline, one
needs at fixed rotation rate to have a larger latitudinal entropy variation or equivalently a thinner tachocline.
Indications of the nature of profiles should be obtained from asteroseismology, in the next years. Even although the quality of asteroseismic data will not be comparable to that available for helioseismology, reliable indications for rotation rate could be derived with asteroseismology \citep*{GizonS03,BallotG06}, and the inner differential rotation could even be inferred for rapid rotators \citep{GizonS04}.

\section{Conclusions and perspectives}\label{sec:concl}

Our simplified 3-D numerical simulations of the turbulent convective envelope in young solar-like stars have served to catch some aspects of the very complex dynamics which occurs in such thick shells and show us how convection and rotation are intricate, even if we are, for sure, far from a true young star.
We have especially seen how these simulations are sensitive to the Taylor number $T_a$.

The main important result of this work concerns the behavior of mean flows according to the rotation speed of the star. We find that the contrast of the differential rotation decreases only slightly, whereas the intensity of the meridional circulation clearly decays.
High $T_a$ make our rotation profiles of rapidly rotating envelopes highly cylindrical.
It is possible to make them more conical ---as exhibited by the true Sun--- by imposing at the bottom of the convective zone a thermal forcing in latitude, such as may arise from a sharp stellar tachocline.

While global convective patterns are quite similar to those seen in solar simulations, some clear differences have appeared. For lower $P_r$ hot spots emerge (see Yc1). Such spots are also seen in a thin solar envelope, but with higher rotation rates \citep{BrownB07}. At higher rotation rates, hot spots change into localized convection, where convective motions are preferentially developed in a particular range of longitudes in the equatorial zone.
Oscillations in convection appear in models characterized by high $T_a$. The shell thickness has to play a role, by favoring the appearance of a strong-shear layer in the middle of the shell. The situation is similar to those also observed in geophysics simulations \citep{GroteB01}.
The mechanism of oscillations is well understood, especially the role of the Reynolds stresses that drive a strong differential rotation which sparks off the shear.
If this kind of vacillating and/or localized deep convection occurs in real stars, their surfaces should be affected. It could be seen in observations --- of luminosity, line widths... --- as temporal variations, arising from changing in the convective state and/or from the rotation of different longitudes into the line of sight.
Young stars, especially T~Tauri stars, are known to be variable objects, but the luminosity fluctuations are linked to magnetic field (chromospheric or coronal activity, photospheric spots...) or, in some cases, accretion phenomena. In the present situation, it would be specious to make any comparison of the deep-convection behavior seen in our most rapidly rotating simulations, with  observed stellar luminosity curves (dominated by surface effects). Nevertheless, if we focus on the differential rotation variability seen in simulations, we note that significant secular variations of differential rotation contrast have also been observed in young K dwarfs with high rotation rate like AB Dor \citep*{CameronD02,DonatiC03a}. However, these observed variations can also have a magnetic origin.

The next step of this work would be to see the effect of these specific properties both of convection and of mean flows, on the dynamo processes which can occur in these stars.
These purely hydrodynamical simulations have exhibited obvious differences with the solar ones, that let forshadow also differences with the Sun, when the magnetic field is included.
Observations tend to show that magnetic properties of young stars are peculiar. All of them show a magnetic activity which can be tracked
by their X-ray emissions. Several issues have been addressed about the signification of correlations ---or the lack of correlation--- between rotation rate and magnetic activity. Is it just an extension of the saturation \citep{StaufferC94,Randich97} and ``supersaturation" phenomena \citep{ProsserR96,James00} observed for main sequence stars? Or is it the signature of a turbulent dynamo distributed throughout the deep convective zone \citep{DurneyD93} instead of
a classic $\alpha$-$\Omega$ dynamo \citep{Parker93}, as suggested by \citet{FeigelsonG03}? Zeeman-Doppler imaging of young
fast rotators \citep{DonatiC03b} shows large-scale azimuthal magnetic structures, that
seems to plead for such dynamo mechanisms.
Such fascinating issues encourage us to pursue simulations of young stars by including magnetic field, and searching for clues to the answers of these, and other, questions.

\acknowledgments
First we acknowledge useful discussions with J. Toomre and B. Brown. We also
wish to thank the referee, P. Gilman, for helping to clarify the paper.
The simulations were performed using the computer facilities of CEA/CCRT.
J.B. is grateful to the developers of the ASH code for letting him use it for this study.

\clearpage

\begin{deluxetable}{ccccccccccc}
\tabletypesize{\scriptsize}
\tablecaption{Parameters and characteristics of models\label{Tab:descrip}}
\tablehead{
\colhead{Model} & \colhead{$\Omega/\Omega_\sun$} & \colhead{$\nu$} & 
\colhead{$\kappa$} & \colhead{$P_r$} &
\colhead{$R_a$} & \colhead{$T_a$} & \colhead{$R_{oc}$} & \colhead{$R_o$} & \colhead{$R_{e}/R_{e}'$} &
\colhead{$N_r,N_\theta,N_\phi$}}
\startdata
Ya1\dotfill & 1 & $3.7\times 10^{12}$ & $9.3\times 10^{11}$ & 4 & $1.7\times 10^{5}$ & 
$1.8\times 10^{6}$ & $0.15$ & $0.013$ & 22/18 & $65,128,256$\\
Ya2\dotfill & 2 & $3.7\times 10^{12}$ & $9.3\times 10^{11}$ & 4 & $3.8\times 10^{5}$ & 
$7.2\times 10^{6}$ & $0.11$ & $0.0053$ & 18/15 & $65,128,256$\\
Ya5\dotfill & 5 & $3.7\times 10^{12}$ & $9.3\times 10^{11}$ & 4 & $1.1\times 10^{6}$ & 
$4.5\times 10^{7}$ & $0.08$ & $0.0016$ & 13/11 & $65,128,256$\\
Ya5T\dotfill & 5 & $9.1\times 10^{11}$ & $2.3\times 10^{11}$ & 4 & $1.2\times 10^{7}$& 
$7.5\times 10^{8}$ & $0.06$ & $0.0020$ & 119/54 & $129,256,512$\\
Yb1\dotfill & 1 & $1.9\times 10^{12}$ & $1.9\times 10^{12}$ & 1 & $4.0\times 10^5$ &
$7.2\times 10^6$ & $0.23$ & $0.014$ & 83/38 & $65,128,256$ \\
Yb2\dotfill & 2 & $1.9\times 10^{12}$ & $1.9\times 10^{12}$ & 1 & $8.6\times 10^5$ &
$2.9\times 10^7$ & $0.17$ & $0.0059$ & 72/31 & $65,128,256$ \\
Yb5\dotfill & 5 & $1.9\times 10^{12}$ & $1.9\times 10^{12}$ & 1 & $2.4\times 10^6$ &
$1.8\times 10^8$ & $0.11$ & $0.0020$ & 46/27 & $65,128,256$ \\
Yb5T\dotfill & 5 & $9.3\times 10^{11}$ & $9.3\times 10^{11}$ & 1 & $7.4\times 10^6$ &
$7.5\times 10^8$ & $0.10$ & $0.0020$ & 155/55 & $129,256,512$\\
Yc1\dotfill & 1 & $4.6\times 10^{11}$ & $1.8\times 10^{12}$ & $1/4$ & $2.7\times 10^6$ &
$1.2\times 10^8$ & $0.29$ & $0.020$ & 628/224 & $129,256,512$ \\
Yc2\dotfill & 2 & $4.6\times 10^{11}$ & $1.8\times 10^{12}$ & $1/4$ & $5.0\times 10^6$ &
$4.8\times 10^8$ & $0.20$ & $0.0084$ & 557/180 & $129,256,512$ \\
Yc5\dotfill & 5 & $4.6\times 10^{11}$ & $1.8\times 10^{12}$ & $1/4$ & $1.1\times 10^7$ &
$3.0\times 10^9$ & $0.12$ &  $0.0024$\tablenotemark{a} & 490/128\tablenotemark{a} & $129,256,512$ \\
Yc5S\tablenotemark{\dag}\dotfill & 5 & $4.6\times 10^{11}$ & $1.8\times 10^{12}$ & $1/4$ &$1.1\times 10^7$ &
$2.9\times 10^9$ & $0.12$ & $0.0025$\tablenotemark{a} & 728/138\tablenotemark{a} & $129,256,512$ \\
Yc5T\dotfill & 5 & $2.3\times 10^{11}$ & $9.3\times 10^{11}$ & $1/4$ & $3.3\times 10^7$ &
$1.2\times 10^{10}$ & $0.11$ & $0.0024$\tablenotemark{a} & 1253/256\tablenotemark{a} & $161,512,1024$ \\
Yd5\dotfill & 5 & $2.3\times 10^{11}$ & $1.8\times 10^{12}$ & $1/8$ & $2.1\times 10^7$ &
$1.2\times 10^{10}$ &$0.12$ & $0.0032$\tablenotemark{a} & 1297/345\tablenotemark{a} & $161,512,1024$ \\
\enddata
\tablecomments{All simulations have an inner radius $r_b=4.2\times10^{10}\unit{cm}$
and an outer radius $r_t=7.3\times10^{10}\unit{cm}$, what is to say 
a thickness of the computational domain $L=3.1\times10^{10}\unit{cm}$.
The eddy viscosity $\nu$ and conductivity $\kappa$ at mid-depth are quoted in \unit{cm^2\,s^{-1}}.
Here evaluated at mid-layer depth are the temporal average of the Prandtl number $P_r=\nu/\kappa$, 
the Rayleigh number $R_a=-(d\Sbb/dr)(\partial\rho/\partial S)gL^4/(\rb\nu\kappa)$,
the Taylor number $T_a=4\Omega_o^2L^4/\nu^2$,
the convective Rossby number $R_{oc}=[R_a/(T_aP_r)]^{1/2}$, 
the rms Rossby number $R_o=\Vvr'/(2\Omega_oL)$,
and the rms Reynolds numbers $R_e=\Vvr L/\nu$ and $R_e'=\Vvr'L/\nu$, where $\Vvr$ and $\Vvr'$ are the rms velocity and rms convective velocity at
middepth (cf. Table~\ref{Tab:veloc}).
The number of radial, latitudinal and longitudinal mesh points are $N_r$, $N_\theta$, $N_\phi$.
}
\tablenotetext{\dag}{Model with a thermal forcing (see \S~\ref{ssec:thforcing}).}
\tablenotetext{a}{Due to vacillating behavior of convection in these models, rms velocities fluctuate noticeably, thus $R_o$ and $R_e$ vary a lot according to the convective level.}
\end{deluxetable}

\begin{deluxetable}{cccccccccccccc}
\tabletypesize{\scriptsize}
\tablewidth{0pt}
\rotate
\tablecaption{Representative velocities, energy contents, differential rotation, and temperature contrast\label{Tab:veloc}}
\tablehead{
\colhead{Model} & \colhead{$\Vrr$} & \colhead{$\Vtr$} & \colhead{$\Vpr$} & \colhead{$\Vpr'$} & \colhead{$\Vvr$} & \colhead{$\Vvr'$} & \colhead{$a_r$} & 
\colhead{KE} & \colhead{DRKE} & \colhead{CKE} & \colhead{MCKE} & \colhead{$\Delta\Omega_\mathrm{top}$} & \colhead{$\Delta T_\mathrm{bot}$}}
\startdata
Ya1\dotfill & 15 & 11 & 18 & 10 & 26 & 21 &0.48 & $1.7\times 10^{6}$ & $6.1\times 10^{5}$ (36\%)& $1.1\times 10^{6}$ (64\%)& 
$3.5\times 10^{3}$ (0.20\%)& 22 & 1.9\\
Ya2\dotfill & 12 & 9 & 15 & 8 & 21 & 17 & 0.51 & $1.2\times 10^{6}$ & $4.4\times 10^{5}$ (37\%)& $7.4\times 10^{5}$ (63\%) & 
$1.2\times 10^{3}$ (0.10\%)& 17 & 3.2\\
Ya5\dotfill & 9 & 7 & 10 & 6 & 15 & 13 & 0.54 & $5.5\times 10^{5}$ & $1.7\times 10^{5}$ (32\%)& $3.7\times 10^{5}$ (68\%)& 
$2.5\times 10^{2}$ (0.045\%) & 10 & 5.4\\
Ya5T\dotfill & 11 & 8 & 32 & 8 & 35 & 16 & 0.47 & $4.3\times 10^{6}$ & $3.7\times 10^{6}$ (86\%) & $6.2\times 10^{5}$  (14\%)& 
$5.0\times 10^{2}$ (0.012\%) & 52 & 11.0\\
Yb1\dotfill & 15 & 12 & 46 & 13 & 50 & 23 & 0.41 & $7.1\times 10^{6}$ & $5.7\times 10^{6}$ (80\%) & $1.4\times 10^{6}$ (20\%) & 
$5.9\times 10^{3}$ (0.083\%) & 64 & 3.8\\
Yb2\dotfill & 12 & 9 & 40 & 11 & 43 & 19 & 0.41 & $5.4\times 10^{6}$ & $4.6\times 10^{6}$ (85\%) & $8.3\times 10^{5}$ (15\%) & 
$1.4\times 10^{3}$ (0.026\%) & 53 & 5.6\\
Yb5\dotfill & 10 & 7 & 24 & 10 & 28 & 16 & 0.42 & $2.2\times 10^{6}$ & $1.7\times 10^{6}$ (78\%) & $4.7\times 10^{5}$ (22\%) & 
$8.1\times 10^{1}$ (0.004\%) & 29 & 6.0\\
Yb5T\dotfill & 10 & 8 & 44 & 10 & 45 & 16 & 0.42 & $7.6\times 10^{6}$ & $7.0\times 10^{6}$ (92\%) & $6.0\times 10^{5}$ (8\%) & 
$4.7\times 10^{2}$ (0.006\%) & 64 & 12.6\\
Yc1\dotfill & 19 & 20 & 89 & 20 & 93 & 33 & 0.30 & $2.9\times 10^{7}$ & $2.6\times 10^{7}$ (89\%) & $3.0\times 10^{6}$ (11\%) & 
$2.0\times 10^{4}$ (0.070\%) & 139 & 4.9\\
Yc2\dotfill & 15 & 14 & 86 & 16 & 88 & 27 & 0.33 & $2.8\times 10^{7}$ & $2.6\times 10^{7}$ (93\%) & $1.9\times 10^{6}$ (7\%) & 
$9.9\times 10^{3}$ (0.036\%)  & 129 & 8.8\\
Yc5\dotfill & 12 & 9 & 71 & 11 & 72 & 19 & 0.40 & $1.8\times 10^{7}$ & $1.7\times 10^{7}$ (95\%) & $9.6\times 10^{5}$ (5\%) & 
$2.3\times 10^{3}$ (0.013\%)& 103 & 12.7\\
Yc5S\dotfill & 10 & 12 & 107 & 12 & 108 & 21 & 0.23 & $3.5\times 10^{7}$ & $3.3\times 10^{7}$ (96\%) & $1.5\times 10^{6}$ (4\%) & 
$3.4\times 10^{3}$ (0.010\%) & 127 & 45.0\\
Yc5T\dotfill & 11 & 10 & 92 & 11 & 93 & 19 & 0.36 & $3.3\times 10^{7}$ & $3.2\times 10^{7}$ (97\%) & $9.9\times 10^{5}$ (3\%)& 
$2.8\times 10^{3}$ (0.009\%) & 140 & 15.4\\
Yd5\dotfill & 15 & 14 & 94 & 15 & 96 & 26 & 0.34 & $3.1\times 10^{7}$ & $2.9\times 10^{7}$ (95\%) & $1.7\times 10^{6}$ (5\%)& 
$8.2\times 10^{3}$ (0.027\%) & 130 & 15.4\\
\enddata
\tablecomments{Temporal averages of the rms velocity $\Vvr$, of the rms components $\Vrr$, $\Vtr$, $\Vpr$, and of
fluctuating velocities  $\Vvr'$ and $\Vpr'$ (axisymmetric components are removed) are 
estimated at midlayer depth, all expressed in units of \unit{m\,s^{-1}}.  $a_r=\Vrr'{}^2/\Vvr'{}^2$ is the anisotropy index. Also listed
are the time average over the complete volume of the total kinetic energy KE
and that associated with the (axisymmetric) differential rotation DRKE, the
(axisymmetric) meridional circulation MCKE, and the non-axisymmetric convection CKE,
all in units of \unit{erg\,cm^{-3}}. The latitudinal contrasts between 0\degr\ and 60\degr\ of angular frequencies $\Delta\Omega$ (resp. temperatures $T$), quoted in nHz (resp. K), are computed at the top (resp. bottom) of the domain.}
\end{deluxetable}

\clearpage

\begin{thebibliography}{67}

\bibitem[{{Adelberger} {et~al.}(1998){Adelberger}, {Austin}, {Bahcall},
  {Balantekin}, {Bogaert}, {Brown}, {Buchmann}, {Cecil}, {Champagne}, {de
  Braeckeleer}, {Duba}, {Elliott}, {Freedman}, {Gai}, {Goldring}, {Gould},
  {Gruzinov}, {Haxton}, {Heeger}, {Henley}, {Johnson}, {Kamionkowski},
  {Kavanagh}, {Koonin}, {Kubodera}, {Langanke}, {Motobayashi}, {Pandharipande},
  {Parker}, {Robertson}, {Rolfs}, {Sawyer}, {Shaviv}, {Shoppa}, {Snover},
  {Swanson}, {Tribble}, {Turck-Chi{\` e}ze}, \& {Wilkerson}}]{Adelberger98}
{Adelberger}, E.~G., {Austin}, S.~M., {Bahcall}, J.~N., {et~al.} 1998, Reviews
  of Modern Physics, 70, 1265

\bibitem[{{Alexander} \& {Ferguson}(1994)}]{Alexander94}
{Alexander}, D.~R. \& {Ferguson}, J.~W. 1994, \apj, 437, 879

\bibitem[{{Ballot} {et~al.}(2006){Ballot}, {Garc{\'{\i}}a}, \&
  {Lambert}}]{BallotG06}
{Ballot}, J., {Garc{\'{\i}}a}, R.~A., \& {Lambert}, P. 2006, \mnras, 369, 1281

\bibitem[{{Ballot} {et~al.}(2004){Ballot}, {Turck-Chi{\` e}ze}, \&
  {Garc{\'{\i}}a}}]{Ballot04BCZ}
{Ballot}, J., {Turck-Chi{\` e}ze}, S., \& {Garc{\'{\i}}a}, R.~A. 2004, \aap,
  423, 1051

\bibitem[{{Barnes} {et~al.}(2005){Barnes}, {Cameron}, {Donati}, {James},
  {Marsden}, \& {Petit}}]{BarnesC05}
{Barnes}, J.~R., {Cameron}, A.~C., {Donati}, J.-F., {et~al.} 2005, \mnras, 357,
  L1

\bibitem[{{Bouvier} {et~al.}(1997){Bouvier}, {Forestini}, \&
  {Allain}}]{BouvierF97}
{Bouvier}, J., {Forestini}, M., \& {Allain}, S. 1997, \aap, 326, 1023

\bibitem[{{Brown} {et~al.}(2007){Brown}, {Browning}, {Brun}, \&
  {Toomre}}]{BrownB07}
{Brown}, B.~P., {Browning}, M.~K., {Brun}, A.~S., \& {Toomre}, J. 2007, \apj, submitted

\bibitem[{{Browning} {et~al.}(2004){Browning}, {Brun}, \&
  {Toomre}}]{BrowningB04}
{Browning}, M.~K., {Brun}, A.~S., \& {Toomre}, J. 2004, \apj, 601, 512

\bibitem[{{Brun}(2007)}]{Brun07}
{Brun}, A.~S. 2007, Astronomische Nachrichten, 328, 329

\bibitem[{{Brun} {et~al.}(2005){Brun}, {Browning}, \& {Toomre}}]{BrunB05}
{Brun}, A.~S., {Browning}, M.~K., \& {Toomre}, J. 2005, \apj, 629, 461

\bibitem[{{Brun} {et~al.}(2004){Brun}, {Miesch}, \& {Toomre}}]{BrunM04}
{Brun}, A.~S., {Miesch}, M.~S., \& {Toomre}, J. 2004, \apj, 614, 1073

\bibitem[{{Brun} \& {Toomre}(2002)}]{BrunT02}
{Brun}, A.~S. \& {Toomre}, J. 2002, \apj, 570, 865

\bibitem[{{Brun} {et~al.}(1999){Brun}, {Turck-Chi{\` e}ze}, \&
  {Zahn}}]{BrunT99}
{Brun}, A.~S., {Turck-Chi{\` e}ze}, S., \& {Zahn}, J.~P. 1999, \apj, 525, 1032

\bibitem[{{Busse}(2002)}]{Busse02}
{Busse}, F.~H. 2002, Physics of Fluids, 14, 1301

\bibitem[{{Cameron} {et~al.}(2001){Cameron}, {Barnes}, {Kitchatinov}, \&
  {Donati}}]{CameronB01}
{Cameron}, A.~C., {Barnes}, J.~R., {Kitchatinov}, L., \& {Donati}, J.-F. 2001,
  in ASP Conf. Ser. 223, 11th Cambridge Workshop on Cool Stars, Stellar Systems
  and the Sun, ed. R.~J. {Garcia Lopez} \& R.~{Rebolo} (San Francisco:
  Astronomical Society of the Pacific), 251

\bibitem[Cameron \& Donati(2002)]{CameronD02} 
Cameron, A.~C., \& Donati, J.-F.\ 2002, \mnras, 329, L23 

\bibitem[{{Chandrasekhar}(1961)}]{Chandrasekhar}
{Chandrasekhar}, S. 1961, {Hydrodynamic and hydromagnetic stability}
  (International Series of Monographs on Physics, Oxford: Clarendon)

\bibitem[{{Clune} {et~al.}(1999){Clune}, {Elliott}, {Glatzmaier}, {Miesch}, \&
  {Toomre}}]{CluneE99}
{Clune}, T.~C., {Elliott}, J.~R., {Glatzmaier}, G.~A., {Miesch}, M.~S., \&
  {Toomre}, J. 1999, Parallel Computing, 25, 361

\bibitem[{{Dobler} {et~al.}(2006){Dobler}, {Stix}, \&
  {Brandenburg}}]{DoblerS06}
{Dobler}, W., {Stix}, M., \& {Brandenburg}, A. 2006, \apj, 638, 336

\bibitem[{{Donahue} {et~al.}(1996){Donahue}, {Saar}, \&
  {Baliunas}}]{DonahueS96}
{Donahue}, R.~A., {Saar}, S.~H., \& {Baliunas}, S.~L. 1996, \apj, 466, 384

\bibitem[{{Donati} {et al.}(2003a){Donati}, {Cameron}, \& {Petit}}]{DonatiC03a}
Donati, J.-F.,  Cameron, A.~C., \& Petit, P.\ 2003a, \mnras, 345, 1187 

\bibitem[{{Donati} {et~al.}(2003b){Donati}, {Cameron}, {Semel},
  {Hussain}, {Petit}, {Carter}, {Marsden}, {Mengel}, {L{\' o}pez Ariste},
  {Jeffers}, \& {Rees}}]{DonatiC03b}
{Donati}, J.-F., {Cameron}, A.~C., {Semel}, M., {et~al.} 2003b, \mnras,
  345, 1145

\bibitem[{{Durney} {et~al.}(1993){Durney}, {De Young}, \&
  {Roxburgh}}]{DurneyD93}
{Durney}, B.~R., {De Young}, D.~S., \& {Roxburgh}, I.~W. 1993, \solphys, 145,
  207

\bibitem[{{Elliott} {et~al.}(2000){Elliott}, {Miesch}, \&
  {Toomre}}]{ElliottM00}
{Elliott}, J.~R., {Miesch}, M.~S., \& {Toomre}, J. 2000, \apj, 533, 546

\bibitem[{{Feigelson} {et~al.}(2003){Feigelson}, {Gaffney}, {Garmire},
  {Hillenbrand}, \& {Townsley}}]{FeigelsonG03}
{Feigelson}, E.~D., {Gaffney}, J.~A., {Garmire}, G., {Hillenbrand}, L.~A., \&
  {Townsley}, L. 2003, \apj, 584, 911

\bibitem[{{Gilman}(1977)}]{Gilman77}
{Gilman}, P.~A. 1977, Geophys. Astrophys. Fluid Dynamics, 8, 93

\bibitem[{{Gizon} \& {Solanki}(2003)}]{GizonS03}
{Gizon}, L. \& {Solanki}, S.~K. 2003, \apj, 589, 1009

\bibitem[{{Gizon} \& {Solanki}(2004)}]{GizonS04}
{Gizon}, L. \& {Solanki}, S.~K. 2004, \solphys, 220, 169

\bibitem[{{Glatzmaier} \& {Gilman}(1981)}]{GlatzGil81b}
{Glatzmaier}, G.~A. \& {Gilman}, P.~A. 1981, \apjs, 45, 351

\bibitem[{{Grote} \& {Busse}(2001)}]{GroteB01}
{Grote}, E. \& {Busse}, F.~H. 2001, Fluid Dynamics Research, 28, 349

\bibitem[{{Henry} {et~al.}(1995){Henry}, {Eaton}, {Hamer}, \&
  {Hall}}]{HenryE95}
{Henry}, G.~W., {Eaton}, J.~A., {Hamer}, J., \& {Hall}, D.~S. 1995, \apjs, 97,
  513

\bibitem[{{Hurlburt} {et~al.}(1986){Hurlburt}, {Toomre}, \&
  {Massaguer}}]{HurlburtT86}
{Hurlburt}, N.~E., {Toomre}, J., \& {Massaguer}, J.~M. 1986, \apj, 311, 563

\bibitem[{{Iglesias} \& {Rogers}(1996)}]{OPAL96}
{Iglesias}, C.~A. \& {Rogers}, F.~J. 1996, \apj, 464, 943

\bibitem[{{James} {et~al.}(2000){James}, {Jardine}, {Jeffries}, {Randich},
  {Cameron}, \& {Ferreira}}]{James00}
{James}, D.~J., {Jardine}, M.~M., {Jeffries}, R.~D., {et~al.} 2000, \mnras,
  318, 1217

\bibitem[{{Jouve} \& {Brun}(2007)}]{JouveB07}
{Jouve}, L. \& {Brun}, A.~S. 2007, \aap, accepted

\bibitem[{{Kitchatinov} \& {R{\"u}diger}(1995)}]{KitchatinovR95}
{Kitchatinov}, L.~L. \& {R{\"u}diger}, G. 1995, \aap, 299, 446

\bibitem[{{Kitchatinov} \& {R{\"u}diger}(1999)}]{KitchatinovR99}
{Kitchatinov}, L.~L. \& {R{\"u}diger}, G. 1999, \aap, 344, 911

\bibitem[{{K{\"u}ker} \& {R{\"u}diger}(1997)}]{KukerR97}
{K{\"u}ker}, M. \& {R{\"u}diger}, G. 1997, \aap, 328, 253

\bibitem[{{K{\"u}ker} \& {Stix}(2001)}]{KukerS01}
{K{\"u}ker}, M. \& {Stix}, M. 2001, \aap, 366, 668

\bibitem[{{Mazumdar} \& {Antia}(2001)}]{MazumdarA01}
{Mazumdar}, A. \& {Antia}, H.~M. 2001, \aap, 368, L8

\bibitem[{{Messina} \& {Guinan}(2003)}]{MessinaG03}
{Messina}, S. \& {Guinan}, E.~F. 2003, \aap, 409, 1017

\bibitem[{{Miesch}(2005)}]{Miesch05}
{Miesch}, M.~S. 2005, Living Reviews in Solar Physics, 2, 1

\bibitem[{{Miesch} {et~al.}(2006){Miesch}, {Brun}, \& {Toomre}}]{MieschB06}
{Miesch}, M.~S., {Brun}, A.~S., \& {Toomre}, J. 2006, \apj, 641, 618

\bibitem[{{Miesch} {et~al.}(2000){Miesch}, {Elliott}, {Toomre}, {Clune},
  {Glatzmaier}, \& {Gilman}}]{MieschE00}
{Miesch}, M.~S., {Elliott}, J.~R., {Toomre}, J., {et~al.} 2000, \apj, 532, 593

\bibitem[{{Monteiro} {et~al.}(2000){Monteiro}, {Christensen-Dalsgaard}, \&
  {Thompson}}]{Monteiro00}
{Monteiro}, M.~J.~P.~F.~G., {Christensen-Dalsgaard}, J., \& {Thompson}, M.~J.
  2000, \mnras, 316, 165

\bibitem[{{Morel}(1997)}]{Morel97}
{Morel}, P. 1997, \aaps, 124, 597

\bibitem[{{Morin} \& {Dormy}(2004)}]{MorinD04}
{Morin}, V. \& {Dormy}, E. 2004, Physics of Fluids, 16, 1603

\bibitem[{{Parker}(1993)}]{Parker93}
{Parker}, E.~N. 1993, \apj, 408, 707

\bibitem[{{Pedlosky}(1987)}]{Pedlosky}
{Pedlosky}, J. 1987, {Geophysical Fluid Dynamics} (New York: Springer)

\bibitem[{{Petit} {et~al.}(2002){Petit}, {Donati}, \& {Cameron}}]{PetitD02}
{Petit}, P., {Donati}, J.-F., \& {Cameron}, A.~C. 2002, \mnras, 334, 374

\bibitem[{{Piau} {et~al.}(2005){Piau}, {Ballot}, \&
  {Turck-Chi{\`e}ze}}]{PiauB05}
{Piau}, L., {Ballot}, J., \& {Turck-Chi{\`e}ze}, S. 2005, \aap, 430, 571

\bibitem[{{Piau} \& {Turck-Chi{\` e}ze}(2002)}]{PiauTC02}
{Piau}, L. \& {Turck-Chi{\` e}ze}, S. 2002, \apj, 566, 419

\bibitem[{{Prosser} {et~al.}(1996){Prosser}, {Randich}, {Stauffer}, {Schmitt},
  \& {Simon}}]{ProsserR96}
{Prosser}, C.~F., {Randich}, S., {Stauffer}, J.~R., {Schmitt}, J.~H.~M.~M., \&
  {Simon}, T. 1996, \aj, 112, 1570

\bibitem[{{Randich}(1997)}]{Randich97}
{Randich}, S. 1997, Memorie della Societa Astronomica Italiana, 68, 971

\bibitem[{{Reiners} \& {Schmitt}(2003)}]{ReinersS03a}
{Reiners}, A. \& {Schmitt}, J.~H.~M.~M. 2003, \aap, 398, 647

\bibitem[{{Rempel}(2005a)}]{Rempel05a}
{Rempel}, M. 2005a, \apj, 622, 1320

\bibitem[{{Rempel}(2005b)}]{Rempel05b}
{Rempel}, M. 2005b, \apj, 631, 1286

\bibitem[{{Rieutord}(2006)}]{Rieutord06}
{Rieutord}, M. 2006, \aap, 451, 1025

\bibitem[{{Rogers}(2000)}]{EOS_OPAL2001a}
{Rogers}, F.~J. 2000, Physics of Plasmas, 7, 51

\bibitem[{{Rogers} {et~al.}(1996){Rogers}, {Swenson}, \&
  {Iglesias}}]{EOS_OPAL96}
{Rogers}, F.~J., {Swenson}, F.~J., \& {Iglesias}, C.~A. 1996, \apj, 456, 902

\bibitem[{{Roxburgh} \& {Vorontsov}(2001)}]{RoxVor01}
{Roxburgh}, I.~W. \& {Vorontsov}, S.~V. 2001, \mnras, 322, 85

\bibitem[{{R{\"u}diger} {et~al.}(1998){R{\"u}diger}, {von Rekowski}, {Donahue},
  \& {Baliunas}}]{RudigervR98}
{R{\"u}diger}, G., {von Rekowski}, B., {Donahue}, R.~A., \& {Baliunas}, S.~L.
  1998, \apj, 494, 691

\bibitem[{{Sestito} {et~al.}(2006){Sestito}, {Degl'Innocenti}, {Prada Moroni},
  \& {Randich}}]{SestitoD06}
{Sestito}, P., {Degl'Innocenti}, S., {Prada Moroni}, P.~G., \& {Randich}, S.
  2006, \aap, 454, 311

\bibitem[{{Skumanich}(1972)}]{Skumanich72}
{Skumanich}, A. 1972, \apj, 171, 565

\bibitem[{{Stauffer} {et~al.}(1994){Stauffer}, {Caillault}, {Gagne}, {Prosser},
  \& {Hartmann}}]{StaufferC94}
{Stauffer}, J.~R., {Caillault}, J.-P., {Gagne}, M., {Prosser}, C.~F., \&
  {Hartmann}, L.~W. 1994, \apjs, 91, 625

\bibitem[{{Ventura} {et~al.}(1998){Ventura}, {Zeppieri}, {Mazzitelli}, \&
  {D'Antona}}]{VenturaZ98}
{Ventura}, P., {Zeppieri}, A., {Mazzitelli}, I., \& {D'Antona}, F. 1998, \aap,
  331, 1011

\bibitem[{{Zahn}(1992)}]{Zahn92}
{Zahn}, J.-P. 1992, \aap, 265, 115

\end{thebibliography}
\end{document}